\documentclass[fleqn,usenatbib]{mnras}

\usepackage{newtxtext,newtxmath}
\usepackage[T1]{fontenc}

\DeclareRobustCommand{\VAN}[3]{#2}
\let\VANthebibliography\thebibliography
\def\thebibliography{\DeclareRobustCommand{\VAN}[3]{##3}\VANthebibliography}

\usepackage{graphicx}	% Including figure files
\usepackage{amsmath}	% Advanced maths commands
\usepackage{multirow}
\usepackage{caption}
\usepackage[dvipsnames]{xcolor}
\usepackage{braket}

\usepackage{graphicx}	% Including figure files
\usepackage{siunitx}
\usepackage{anyfontsize}
\usepackage{subcaption} %Subfigures
\usepackage{makecell} % 
\usepackage{multirow} %
\usepackage{booktabs}

\title[Fundamental Plane of simulated ETGs]{Reconciling the Fundamental Plane of Early-Type Galaxies with hydrodynamical simulations: The case of IllustrisTNG100-1}

\author[de Araujo Ferreira et al.]{
Pedro de Araujo Ferreira,$^{1,\dagger}$\thanks{E-mail: pedroitalo96@academico.ufs.br}
Nicola R. Napolitano$^{2,3, 4,\dagger}$, Crescenzo Tortora$^{4}$, Luciano Casarini$^{1}$, \newauthor 
Francisco Villaescusa-Navarro$^{5,6}$
\\
$^{1}$Physics Department, Federal University of Sergipe, 49100-000, São Cristovão-SE, Brazil;\\
$^{2}${Department of Physics E. Pancini, University Federico II, Via Cinthia 21, I-80126, Naples, Italy};\\
$^{3}$Scuola Superiore Meridionale, Via Mezzocannone, 4, 80138, Naples, Italy;\\
$^{4}${INAF – Osservatorio Astronomico di Capodimonte, Salita Moiariello 16, I-80131 Napoli, Italy};\\
$^5${Center for Computational Astrophysics, Flatiron Institute, 162 5th Avenue, New York, NY, 10010, USA};\\
$^6${Department of Astrophysical Sciences, Princeton University, 4 Ivy Lane, Princeton, NJ 08544 USA};\\
$\dagger$ {These authors contributed equally};
}

\date{Accepted XXX. Received YYY; in original form ZZZ}

\pubyear{\the\year{}}

\begin{document}
\label{firstpage}
\pagerange{\pageref{firstpage}--\pageref{lastpage}}
\maketitle

\begin{abstract}
% Abstract should be less than 200 words
The Fundamental Plane (FP) of Early-Type Galaxies (ETGs) encapsulates a tight correlation among their structural and dynamical properties and provides an important benchmark for galaxy formation models. However, cosmological hydrodynamical simulations have historically struggled to reproduce the observed FP tilt, with discrepancies often attributed to to flawed feedback physics or insufficient resolution. 
Using the IllustrisTNG100-1 simulation, we show that adopting observationally motivated measurements, including Sérsic-derived photometric parameters and dynamically inferred velocity dispersions designed to minimise softening-length effects, substantially reduces the discrepancy between simulated and observed FPs. We further explore the impact of non-universal, mass-dependent Initial Mass Function (IMF) variations through forward modelling of their effects on galaxy structural and dynamical quantities. In particular, bottom-heavy IMF variations produce FP coefficients fully consistent with observational constraints for both direct and orthogonal fits.
Our results suggest that a significant fraction of the long-standing FP tension arises from how galaxy observables are extracted and interpreted in simulations, although residual discrepancies may still reflect limitations in the underlying baryonic physics. These findings highlight the importance of observational realism and IMF variations for interpreting galaxy scaling relations and for improving the predictive power of hydrodynamical simulations of ETG formation.
\end{abstract}
%The Fundamental Plane (FP) of Early-Type Galaxies (ETGs) is an observational relation connecting central velocity dispersion, effective radius, and surface brightness, reflecting their structure and evolution. Hydrodynamical cosmological simulations
%have struggled to reproduce the observed FP, with discrepancies often attributed to incorrect baryonic physics or insufficient numerical resolution. However, mismatches between simulated and observed quantities (the so-called observational realism problem) can significantly contribute to these discrepancies. 
%Here we examine the FP in the IllustrisTNG100‑1 simulation using $\sim$10\,000 ETGs for which we have derived observed‑like parameters: Sérsic‑derived surface brightness and sizes, and a dynamically corrected velocity dispersion minimising resolution effects. Notably, implementing this “observational realism” yields a simulated FP in excellent agreement with observations. Furthermore, by mimicking a non‑universal IMF on the dynamically corrected dispersion brings FP parameters into full observational consistency. This demonstrates, for the first time, that the FP “tilt” in hydrodynamical simulations can be explained without altering feedback recipes, highlighting IMF non‑universality as a crucial missing ingredient in modelling baryonic physics in galaxy formation.


% Main text (introduction/results/discussion) should be less than about 3000 words and should not contain footnotes 
\section{Introduction}
The formation and evolution of Early-Type Galaxies (ETGs) are imprinted in their structural and dynamical properties, which follow well-defined empirical correlations \citep{FaberJackson1976, Kormendy+1977, Jorgensen+1996}. One of the most relevant among these is the so-called Fundamental Plane \citep[FP;][]{Jorgensen+1996, Bernardi_2003b}, a tight relation between effective radius, $R_{\rm e}$, central velocity dispersion, $\sigma_{\rm e} = \sigma(R_{\rm e})$, and central surface brightness, $I_{\rm e} = I(R_{\rm e})$ \citep{Dressler+1987}.
%[$L_{\odot}$/kpc$^2$]) . 
This important correlation not only provides key insights into the physical processes that shape ETGs, from their assembly histories to their dynamical equilibrium \citep{donofrio_2022,2024MNRAS.527..706Z}, but also has several astrophysical applications, 
%as its usefulness 
as, e.g., a distance indicator  \citep{DOnofrio+97_distance}, or to the mapping of the peculiar velocity field of galaxies \citep{Willick+1995, Strauss_1995}.

It is customary to parameterise the FP as (see, e.g., \citealt{Jorgensen+1996,Bernardi_2003b, Hyde+2009, Saulder+2013}):
\begin{equation}
    \log R_{\rm e} = a\log\sigma_{\rm e}
    + b\log \langle I_{\rm e}\rangle + c,
    \label{FP}
\end{equation}
where $a,~b$ and $c$ are free parameters, and  $\langle I_{\rm e}\rangle$ is the mean surface brightness, defined as $\langle I_{\rm e}\rangle = L/(2\pi R_{\rm e}^2)$, where 
%In particular, we used 
the total luminosity $L$ of each ETG is refereed to a specific filter (e.g., the $r$-band, $L_{\rm r}$, in this work). %, and the LOS velocity dispersion $\sigma_{\rm e}$ is defined according to the corrections presented above. 
%The FP encapsulates other classical scaling laws in its various projections, including the Faber-Jackson relation\cite{FaberJackson1976}, which links total luminosity to velocity dispersion ($L_{\rm tot} \propto \sigma_{\rm e}^{\alpha}$, with typically $\alpha \sim 4$), and the Kormendy relation\cite{Kormendy+1977, Bernardi+2003}, which connects surface brightness and effective radius ($\langle I_{\rm e}\rangle \propto R_{\rm e}^{\nu}$, $\nu \sim 1.3$ in the optical band). %For fully virialized ETGs and constant dynamical mass-to-light ratio $M_{\rm dyn}/L$, one should obtain $a=2$ and $b=-1$ \cite{Mo}. However, observations show a considerable ``tilt'', which is usually interpreted in terms of the dependence of the $M_{\rm dyn}/L$ on $M_{\rm dyn}$ or $L$\cite{Cappellari+2006}. In particular, we focus on the validation of the $a$ and $b$ FP parameters, as the $c$ parameter is primarily a normalization factor dependent on a galaxy’s structural properties and dynamics, and it is less central to understanding the physical scaling relations of the FP.
For fully virialized ETGs with a constant dynamical (or total) mass-to-light ratio, $M_{\rm dyn}/L$, the virial theorem predicts an FP with slopes $a = 2$ and $b = -1$ (see, e.g., \citealt{Mo,Binney}), but observations reveal a tilted FP ($a \sim 1.3$, $b \sim -0.75$; e.g., \citealt{Saulder+2013} and references therein), which is often interpreted as a systematic increase in $M_{\rm dyn}/L$ with galaxy mass or luminosity \citep{Cappellari+2006}. This tilt arises from multiple physical factors: stellar population gradients (older, metal-rich stars in massive ETGs elevate $M_{\rm dyn}/L$ via passive evolution;~\citealt{Gallazzi+2005, Graves+2010ApJ}), Initial Mass Function (IMF) variations (e.g., bottom-heavy IMFs in massive galaxies boost the stellar mass-to-light ratio $M_{\star}/L$;~\citealt{Treu+2010}), increasing dark matter fractions with mass \citep{Tortora+2009,Auger+2010}, and structural non-homology (e.g., Sérsic index trends or orbital anisotropy altering the virial coefficient linking the galaxy mass and the $R_{\rm e}$ and $\sigma_{\rm e}$ in the virial theorem, \citealt{Bernardi+2003}). Collectively, these mechanisms reflect how galaxy assembly -- shaped by hierarchical merging, feedback, and dark matter halo coupling \citep{Mo} -- breaks self-similarity, imprinting the tilt as a tracer of mass-dependent physics. While $a$ and $b$ in Eq. (\ref{FP}) encode these dynamical and evolutionary processes, the intercept $c$ is often secondary, tied to calibration rather than intrinsic galaxy properties \citep{Cappellari+2013}. Thus, in our analysis, we will discard the $c$ parameter and focus on characterising $a$ and $b$.
%, as the $c$ parameter is primarily a normalization factor and is less central to understanding the physical scaling relations of the FP.

%Moreover, t
The values of the FP parameters can depend on the photometric band -- with $a$ being more affected than
$b$
%has a very weak dependence
\citep{Bernardi_2003b, Hyde+2009, Saulder+2013}, and the adopted fitting method \citep{Bernardi_2003b, Saulder+2013}. The two most common fitting approaches are the ``direct'' and ``orthogonal'' fits. In the direct fits,  one of the variables (typically the effective radius) is chosen as the dependent variable, and a linear regression is performed on the other two ($\sigma_{\rm e}$ and $\langle I_{\rm e}\rangle$). This method minimises the residuals only in the direction of the chosen dependent variable. It is particularly useful if the goal is to use the FP as a distance indicator (if the dependent variable is chosen to be $R_{\rm e}$). 
On the other hand, the orthogonal fits minimise the residuals in the direction orthogonal to the plane. This method finds the best-fitting plane that minimises the distance from all points to the plane in all three dimensions simultaneously.
%, i.e., it treats all variables symmetrically. 
The latter method provides a more physically interpretable FP in terms of the Virial Theorem \citep{Bernardi_2003b}.  %In particular, our primary analysis is focused on direct fits, as we acknowledge the fact that this current study lacks on a detailed analysis of the error covariance among the FP parameters. We suspect that, if the errors on the virtual-ETG FP variables are strongly correlated, it may substantially impact the orthogonal fits with respect to observations. We plan to assess the impact of these correlated errors in future analyses.
{In our analysis, we present the results obtained with both approaches}.

Although cosmological hydrodynamical simulations reproduce many galaxy properties \citep{Pillepich+2017, Nelson+2018, Tang+2021, ferreira2025cataloguevirtualearlytypegalaxies}, they still fail to match the observed FP self‑consistently. This is 
%particularly true for 
found, e.g., in Illustris‑1 \citep{Vogelsberger+2014}, IllustrisTNG100‑1 \citep{Nelson+2018, Pillepich+2018, Marinacci+2018, Springel+2018, Naiman+2018} and EAGLE simulations \citep{Schaye+2014} -- see. e.g., \citet{D.Xu, Lu+2020, deGraaff+2022, Donofrio_2024}. The discrepancy is often attributed to limited baryonic models, such as feedback, baryon–dark matter ratios, or to resolution \citep[i.e., softening length;][]{Lu+2020, deGraaff+2022}. However, no systematic study has assessed how parameter extraction impacts FP inferences, despite the known “observational realism” problem~\citep{Pillepich+2017, Tang+2021, wu2023total, ferreira2025cataloguevirtualearlytypegalaxies}. In particular, velocity dispersions from simulations~\citep{D.Xu, Y.Wang, Lu+2020, deGraaff+2022} are affected by resolution, as the gravitational potential is artificially shallower than it should be in an ideal case without the introduction of the ``softening length''~\citep{2019_vel_dis_under_hint2, 2020_vel_dis_under_hint1}. Consequently, central velocity dispersions below the softening length are underestimated. In the following, we will refer to this underestimated $\sigma_{\rm e}$ as the “softened” velocity dispersion.

To address some critical aspects of the observational realism problem, in our recent work~\citep{ferreira2025cataloguevirtualearlytypegalaxies}, we have analysed simulated ETGs against observational data by creating a catalogue from the IllustrisTNG100-1 simulation, with more realistic photometric and kinematic features. Observables were derived using \citet{Sersic} profile modelling for photometric parameters and dynamical modelling~\citep{Tortora+2014, Tortora+2022} for velocity dispersion. This ``virtual-ETG" catalogue includes $10\,121$ local systems ($z \leq 0.1$) with stellar masses between $10.47 \leq \log M_\star/M_{\odot} \leq 11.95$, all identified as central ETGs of their groups (see \S\ref{sec:Methods}). We have shown that scaling relations (e.g., size-mass, Kormendy, Faber-Jackson) and the central density profile slopes' properties closely align with observations. In particular, the correlation between velocity dispersion and the total density slope, negative in TNG ETGs but usually positive in observed systems (see, e.g., \citealt{D.Xu, Y.Wang}), has been corrected by redefining the velocity dispersion to minimise effects below the simulation's softening length (see \S\ref{sec:Methods}). 
%These results indicate that previous claims for revisions in baryon physics might have overlooked the effects of simulation resolution on the velocity dispersion. 
To accurately assess the physics in simulations, it is vital to extract parameters that reflect observations and minimise the effects of the spatial resolution.

With that in mind, in this work, we go a step further and test whether the virtual‑ETG catalogue introduced in~\citet{ferreira2025cataloguevirtualearlytypegalaxies}, and specifically the dynamically‑motivated velocity dispersion, can resolve the mismatch between the simulated FP and its observational counterpart, as it has for other scaling relations.

Here we remind that, for the virtual‑ETG sample, a S\'ersic model is used to derive the effective radius $R_{\rm e}$ and mean surface brightness $\langle I_{\rm e}\rangle$, defining the photometric part of the FP. For the kinematic part, we adopt two velocity dispersions: (1) a weighted dispersion $\sigma_{\rm e,corr}$, corrected via statistical weights to mitigate softening‑induced central biases; and (2) a dynamical dispersion $\sigma_{\rm e,dyn}$, obtained by solving the isotropic Jeans equation~\citep{Mo} using the simulated stellar and dark mass profiles, excluding scales below the softening length. This is a practical way to mimic the central dynamics of the virtual-ETGs, eliminating all softening length effects that cannot be avoided in the numerical treatment of the hydro-particles.

 {We generalise the FP parameters to account for IMF variations, as suggested by observations of massive ETGs (e.g., \citealt{Treu+2010, Conroy_van_dokkum2012, Spiniello_2012, TRN13_SPIDER_IMF, Spiniello+2014, Spiniello+2015, Ziegler+2022}). Specifically, we start from the stellar mass, luminosity, and sizes of ETGs extracted from the IllustrisTNG100-1 simulation, which are, by construction, computed assuming a universal Chabrier IMF. We then apply an \textit{a posteriori} IMF prescription to these simulated galaxies by (i) modifying their effective radii to account for the structural changes expected under different IMF assumptions~\citep{Barber_2018}, and (ii) introducing an IMF mismatch factor (\S\ref{sec:Methods}) that rescales the stellar mass and luminosity profiles. This rescaling converts the original Chabrier-based quantities into those corresponding to the non-universal IMF being modelled, allowing us to emulate either bottom-heavy or top-heavy IMF scenarios within the same simulated galaxy population.
%We further generalise $\sigma_{\rm e,dyn}$ to include IMF variations (as indicated by real-world samples of massive ETGs, e.g., refs. \cite{TRN13_SPIDER_IMF, Ziegler+2022}), namely $\sigma_{\rm e, \delta IMF}$ by introducing an IMF mismatch factor (\S\ref{sec:Methods}) to rescale the stellar mass profile via a simplified scaling law proportional to the stellar mass of the virtual-ETGs,  which is based on a universal Chabrier\cite{Chabrier2003} IMF in IllustrisTNG. 
We explored several FP parameterisations under different IMF prescriptions: {(i)} an FP assuming a Chabrier (Chab for short) IMF with the central velocity dispersion given by $\sigma_{\rm e, corr}$, {(ii)} a similar FP formulation but adopting $\sigma_{\rm e, dyn}$ as the dynamical term, {(iii)} an FP constructed with a non-universal Bottom-Heavy (BoH) IMF, and {(iv)} an FP constructed with a non-universal Top-Heavy (ToH) IMF.}

This approach %thus 
brings our analysis into closer alignment with observational methodologies and enhances the interpretability of our results in a realistic astrophysical context.   {We refer the reader to \S\ref{sec:Methods} for further details on these velocity dispersion corrections, S\'ersic fits and non-universal IMF emulations}.

\section{Methods}
\label{sec:Methods}
\subsection{The catalogue}
In this study, we utilise data contained in the virtual-ETG catalogue~\citep{ferreira2025cataloguevirtualearlytypegalaxies}. Below, we describe some of its main aspects and the subset of relevant features. For further details, we refer the reader to~\citet{ferreira2025cataloguevirtualearlytypegalaxies}. This catalogue includes photometric and kinematic properties of synthetic ETGs obtained from the IllustrisTNG100-1 hydrodynamical simulation of galaxy formation~\citep{Nelson+2018, Pillepich+2018, Marinacci+2018, Springel+2018, Naiman+2018}. The simulation consists of a cubic volume with a side length of $110.7$ Mpc, containing baryonic and dark matter mass resolutions of $1.4\times 10^6~M_{\odot}$ and $ 7.5\times 10^6~M_{\odot}$, respectively. The gas softening length is adaptive and has the minimum value $\epsilon_{\rm gas,\min} = 0.185$ kpc. For the stellar and dark matter content, the softening length is fixed in comoving units to $\epsilon_{\rm DM, \star} = 0.74$ kpc at $z< 1$. Furthermore, the TNG100-1 simulation employs the Planck 2015 cosmology~\citep{planck15}, whose density parameters of matter, dark energy, baryons, and the reduced Hubble constant are respectively given by $(\Omega_{\rm m},~\Omega_{\rm \Lambda},~\Omega_{\rm b},~h) = (0.3089,~0.6911,~0.0486,~0.6774)$. The final catalogue consists of $10\,121$ low-redshift ($z\leq 0.1$) virtual-ETGs, whose (projected) stellar masses lie in the range $10.47 \leq \log M_\star/M_{\odot}\leq 11.95$. These galaxies are identified as central objects, i.e., the most massive objects of their respective groups. Below, we summarise the extraction of some key parameters of the catalogue that are of great relevance for this work.

%Properties of the virtual-ETG were determined using observational techniques, such as fitting S\'ersic\cite{Sersic} profiles to characterize the projected light distribution and extracting line-of-sight (LOS) velocity dispersions from dynamical modelling. Further parameters include 3D masses and central density profiles.  

Observational-like parameters in the virtual-ETG catalogue were derived from viewing each ETG along the Cartesian axes ($X,~Y, ~Z$) within the simulation box. %From each axis projection, we extracted their photometric and kinematic information. 
For the extraction of the photometric information, it was assumed that each stellar particle within the galaxies is a Simple Stellar Population (SSP), and used the software \texttt{Flexible Stellar Population Synthesis} \citep{C.Conroy, C.ConroyII, D.Foreman-Mackey} to build synthetic stellar populations given the simulated data of metallicity, birth time and initial mass of the stellar particles. For such, we assumed a \cite{Chabrier2003} IMF and a simple two-component model for dust for attenuation~\citep{S.Charlot}, defined as:
\begin{equation}
     \tau(t) \equiv \begin{cases}
                    1.0\left(\lambda/5500~\mathring{A}\right)^{-0.7}, \quad t\leq 10^7~\text{yr}\\
                    0.3\left(\lambda/5500~\mathring{A}\right)^{-0.7}, \quad t > 10^7~\text{yr}.
                    \end{cases}
\end{equation}
Here, $\tau$ denotes the optical depth. In this framework, the $\tau$ in young stellar systems is linked to the presence of dust within molecular clouds that enclose the young stars. After these molecular clouds are dispersed, typically over $t\sim 10^7$ years \citep{Blitz}, the optical depth corresponds to a consistent screen distributed throughout the galaxy. {As for the dust model, we use the one on \citet{Rodriguez+2019}, which acts similarly to more sophisticated models, as the one also adopted for Illustris and IllustrisTNG, e.g., \citealt{D.Xu, Nelson+2018}, but is computationally lighter. In particular, the two models produce a similar modification of the S\'ersic parameters in the high-mass region where our virtual-ETGs are situated (see \citealt{Rodriguez+2019}, fig. 7).
}
%While previous works on Illustris and IllustrisTNG simulations have employed more advanced dust models (for example, see refs. \cite{D.Xu, Nelson+2018}), we base our modelling on ref. \cite{Rodriguez+2019} (see their fig. 7), and assert that the correlations between our Sérsic-derived parameters and stellar mass remain roughly consistent, irrespective of the dust model employed in their work, particularly at the high-mass region where our virtual-ETGs are situated. This less complex dust modelling also enhances computational efficiency for running multiple SSP fittings on the IllustrisTNG100-1 dataset.

With this light information, the surface brightness profiles for each of the three projections (i.e., along the $X,~Y,$ and $Z$ axes) were extracted and modelled with S\'ersic profiles: 
\begin{equation}
    I(r) = I_{\rm e} \exp\left\{-b_n\left[\left(r/R_{\rm e}\right)^{1/n}-1\right]\right\}, \label{sersic_profile}
    \end{equation} 
where $R_{\rm e}$ is the effective radius, a projected size that encloses half of the total luminosity, $I_{\rm e}$ is the surface brightness at $R_{\rm e}$ measured with respect to the SDSS $r$-band, and $n$ is the S\'ersic index. Lastly, up to fourth order in $\nu=1/n$, the coefficient $b_n$ is given by \cite{MacArthur+2003}:
\begin{equation}
b_n \approx \frac{2}{\nu}-\frac{1}{3}+\frac{4\nu}{405}+\frac{46\nu^2}{25515}+\frac{131\nu^3}{1148175}-\frac{2194697\nu^4}{30690717750}. 
\end{equation}
To mitigate resolution effects in the construction of light profiles, an adaptive partitioning method for fitting was implemented  {and applied for each of the three projections}. Each profile was fitted within a radial range extending from $xR_{\rm e,dir}$ to $3R_{\rm e,dir}$, where $x$ is an adaptive fraction of $R_{\rm e,dir}$. For galaxies with $R_{\rm e,dir} \geq 10$ kpc, $x$ varied between 0.1 and 0.3, while for those with $R_{\rm e,dir} < 10$ kpc, it ranged from 0.3 to 0.4. Here, $R_{\rm e,dir}$ refers to a “direct" effective radius, obtained non-parametrically from the projected light distribution, serving as an initial estimate for the modelled effective radius $R_{\rm e}$. To ensure that the fitted profiles remained above the simulation's softening length, a lower limit of $xR_{\rm e} \geq 1$ kpc was imposed. Finally, multiple profile fits were explored by varying $x$ in five discrete steps within its respective range, and the optimal fit was determined as the one that yielded the smallest $\chi^2$ value. For further details on this adaptive partitioning, we refer the reader to~\citet{ferreira2025cataloguevirtualearlytypegalaxies}.  {As we describe below, the Sérsic profile of each virtual-ETG is then de-projected and multiplied by its stellar mass-to-light ratio to construct the 3D stellar density profiles.}
%We emphasise that the above procedure is equivalent to fitting real galaxies with the observational point‐spread function (PSF) properly incorporated, by convolving the Sérsic model with the PSF before fitting it to the observed light profile (see, e.g., \citenum{2017ApJ...844L...6P}).

For the modelling of the  {DM} density profiles, it was assumed that  {their} central density distributions are reliably described as power-laws~\citep{Tortora+2014, D.Xu, Y.Wang}:
\begin{equation}
 {\rho_{\rm DM}(r) = \rho_{1/2,\rm DM}\left(\frac{r}{r_{1/2}}\right)^{-\gamma_{\rm DM}}}
\end{equation}
where $r_{1/2}$ is the stellar half-mass radius, i.e., the 3D radius enclosing half of the total stellar mass of a given galaxy. %This functional form is assumed for the $i$-th component, which can be either stars, dark matter, gas or the total contribution. 
From the fits, a normalization factor  {$\rho_{1/2,\rm DM}$} [$M_{\odot}$/kpc$^3$] and a logarithmic density slope  {$\gamma_{\rm DM}$} (dimensionless) were extracted. Each profile is modelled over a radial range defined within $r \in [0.4, 4]r_{1/2}$. The lower bound of $0.4r_{1/2}$ ensures that all the profiles are defined above the simulation's softening length. 

In summary, the extraction of parameters associated with light and density profiles was carefully performed to minimise the effects of the simulation's softening length, i.e., these profiles were respectively modelled assuming lower boundaries for the radial ranges always defined above $\epsilon_{\rm DM, \star}$, such that the best fit profiles for each object give a nice analytical extrapolation for scales below $\epsilon_{\rm DM, \star}$.  {The minimum scale is chosen based on the resolution convergence of profiles in both TNG50-1 ($\epsilon_{\rm DM,\star} \approx 0.29$ kpc) and TNG100-1 ($\epsilon_{\rm DM,\star} \approx 0.74$ kpc). In other words, we compare the minimum average radius where both simulations have consistent density profiles (see our Figure \ref{fig:resolution_convergence_comparison}). From the average profiles presented in Figure \ref{fig:resolution_convergence_comparison}, the profiles are reliable above $r \sim 1$ kpc, which is roughly $1.35\epsilon_{\rm DM, \star}$, where $\epsilon_{\rm DM, \star}$ is the softening length of TNG100-1}. 

In this work, to define the photometric part of the FP, we used the following parameters extracted from the S\'ersic modelling:
\begin{itemize}
    %\item S\'ersic index ($n$): This parameter is related to how concentrated the light is in a specific galaxy projection;
    \item Effective radius ($R_{\rm e}$ [kpc]): This is defined as the projected radius enclosing half of a galaxy's total luminosity; 
    \item Total $r$-band luminosity ($L_{\rm r}$ [$L_{\odot}$]): It is determined by integrating the S\'ersic profile over a circular area to infinite extent.
\end{itemize}
With these parameters, we can define a mean central surface brightness $\langle I_{\rm e}\rangle = L_{\rm r}/(2\pi R_{\rm e}^2)$ [$L_{\odot}$/kpc$^2$].

Lastly, to extract the kinematic information, we used a heuristic correction for the velocity dispersion computed using the stellar velocities provided by the simulation to adjust for the absent velocity dispersion beneath $\epsilon_{\rm DM, \star}$, alongside a more reliable method built on the dynamical modelling of ETGs. Below, we offer a brief introduction to these adjustments, with additional details available in the virtual-ETG paper~\citep{ferreira2025cataloguevirtualearlytypegalaxies}.
\begin{itemize}
    \item Weighted LOS velocity dispersion ($\sigma_{\rm e, corr}$ [km/s]): Let {$\sigma_{[1~\mathrm{kpc}, R_{\rm e}]}$} be the LOS velocity dispersion calculated above the TNG100-1 softening length,  {in particular, above $1$ kpc, where the velocity dispersion profiles of TNG100-1 are reliable, as indicated by our Figure~\ref{fig:resolution_convergence_comparison}}. By construction, this velocity dispersion does not suffer from the same biases as the one below $\epsilon_{\rm DM, \star}$. However, by excluding this central contribution, we underestimate the overall $\sigma_{\rm e}$. To adjust for this absent $\sigma$, we multiply  $\sigma_{[1~\mathrm{kpc}, R_{\rm e}]}$ by a statistical weight: 
        \begin{equation}
        \sigma_{\rm e,corr}^2 = \sigma_{[1~\mathrm{kpc}, R_{\rm e}]}^2\left[\frac{\frac{L(1~\mathrm{kpc} < r < R_{\rm e})}{L(<R_{\rm e})}}{\left( 1 - \frac{L(<1~\mathrm{kpc})}{L(<R_{\rm e})} \left(\frac{1~\mathrm{kpc}}{R_{\rm e}}\right)^{-0.132} \right)}\right],
        \label{eq:sig_e_corr}
    \end{equation}  
    where $L(<R)$ represents the luminosity inside each aperture. In particular $L(1~\mathrm{kpc} < r < R_{\rm e}) = L(< R_{\rm e}) - L(< 1~\rm{kpc})$. This correction is stronger for smaller systems and weaker for larger ones.  {This correction is purely heuristic and follows the observational aperture correction with slope $-0.066$~\citep{Cappellari+2006}, resulting in an effective exponent of $-0.132$ after algebraic manipulation (see \citealt{ferreira2025cataloguevirtualearlytypegalaxies}). As this work focuses on a population-level analysis, we adopt this average value for all systems. We acknowledge that this assumption neglects possible effects of non-homologies (e.g., dependence on the Sérsic index, as suggested by \citealt{Zhu_2023}). However, when attempting to test the slopes from this work, we ended up with even weaker corrections, which would render this approach even less effective in producing significant FP corrections. Hence, to maximise the effect of the correction, we conservatively used a routine formula, although this does not capture the full dynamical range imposed by non-homology.
    %which represent a limitation of this correction and will be addressed in future work.
}
    
    \item Dynamical LOS velocity dispersion ($\sigma_{\rm e, dyn}$ [km/s]): Dynamically derived $\sigma_{\rm e, dyn}$ based on the light and mass distribution of each ETG, computed via the spherical Jeans equation (assuming isotropic orbits), projected along the LOS and integrated within an $R_{\rm e}$ aperture, as described in~\cite{Tortora+2022} (eqs. 2--6). %Precisely, for the stellar component, we have used the S\'ersic profile and related quantities, and 
    For the total mass density profile $\rho_{\rm tot}(r)$, we use a two component model given by the sum of the 3D stellar density profile $\rho_{\rm \star}(r)$  {-- which comes from the de-projection of the Sérsic profile --} and the dark matter density profile $\rho_{\rm DM}(r) = \rho_{\rm 0, DM} (r/r_{1/2})^{-\gamma_{\rm DM}}$. We ignore the gas contribution because our simulated systems are consistent with massive ETGs, having very little central gas content~\citep{ferreira2025cataloguevirtualearlytypegalaxies}.  {We stress here that the use of simple assumptions adopted above about spherical symmetry and isotropy is sufficient to show the impact of realistic true dynamics with respect to a softened dynamics, and are not meant to reproduce the intrinsic galaxy motions fully. However, to be more quantitative about the degree of fidelity of these assumptions,
    %Furthermore, it is possible that the assumptions we made for our dynamical modelling may introduce biases in our results. However, follow 
    \cite{Tortora+2014} has shown that adopting a modest radial anisotropy (typical $\beta \sim 0.2$) would change the central line-of-sight velocity dispersion only at the per cent level -- i.e., a few per cent in $\sigma$, corresponding to roughly $1$–$5$ km~s$^{-1}$ depending on the galaxy’s velocity dispersion. Similarly, moderate triaxiality in intermediate-to-high mass ETGs introduces only small systematic variations in ensemble-averaged quantities of galaxy centers. For the population-level analysis presented here, these effects are comparable to, or smaller than, our modelling uncertainties and therefore have only a minor impact on the derived trends~\citep{Cappellari+2020}. Surely, more extreme orbital anisotropies or strongly triaxial shapes can produce larger and non-negligible biases and would require more flexible dynamical modelling -- e.g., Jeans Anisotropic Modelling (JAM) or Schwarzschild -- to control, but should not represent average behaviour in real ETGs.}
 \end{itemize}

\subsection{ {IMF variations in ETGs: context and implementation}}
  {The reason behind the choice of introducing the IMF ``non-universality'' in our analysis is to quantify the \textit{dynamical impact} of the empirically observed IMF-$\sigma$ relation on Fundamental Plane measurements. However, this cannot ignore the long-standing tension in IMF determinations for ETGs; thus, it is useful to place our analysis within the broader astrophysical context. A considerable body of evidence now indicates that the stellar initial mass function (IMF) in massive ETGs is not universal, but instead reflects the physical conditions under which stars formed. Chemical-evolution constraints at high redshift -- notably the high [$\alpha$/Fe] ratios and super-solar metallicities of massive ETGs~\citep{Thomas_2003, Yan_2024} -- favour an early, intense starburst phase characterised by a more top-heavy IMF~\citep{Gunawardhana_2011}. In contrast, dwarf-sensitive absorption features observed in the central regions of present-day ETGs imply an excess of low-mass stars, indicative of a bottom-heavy IMF~\citep{Labarbera2013}, while independent dynamical and lensing analyses often find elevated mass-to-light ratios consistent with a heavier-than-Chabrier IMF~\citep{Cappellari+2013, Tortora_2018MNRAS}. These apparently divergent constraints can be reconciled within a multi-phase, time- and density-dependent IMF framework in which the IMF is top-heavy during the early, high-pressure starburst epoch yet becomes increasingly bottom-heavy in the dense, metal-rich centres assembled at later times. Recent post-processing of hydrodynamical simulations with pressure-dependent IMF prescriptions~\citep{Barber_2018} show that such behaviour naturally emerges from the dependence of the IMF on local interstellar-medium conditions.}
 
Within this broader picture, the %IMF--$\sigma$
 IMF scaling with global galaxy features (such as $M_\star$ or $\sigma$) adopted in our forward-modelling analysis should be regarded as an empirical, end-state parameterisation of these multi-phase processes, rather than as a statement about the instantaneous dependence of the IMF on present-day velocity dispersion. Our approach thus complements, rather than replaces, galaxy-evolution-based IMF studies by isolating how IMF-dependent mass-to-light ratios propagate into the inferred velocity dispersions and the resulting Fundamental Plane coefficients. {In this respect, following \citet{Barber_2018} we have considered two extreme scenarios -- a bottom-heavy and a top-heavy IMF -- to explore the range of variations expected in the FP parameters under different IMF assumptions. This approach is intended to provide an approximate envelope of possible behaviours. However, we caution that more complex IMF variations, including spatial gradients within galaxies \citep{Bernardi+2023} and time evolution \citep{van_Dokkum_2008}, may induce effects that are not fully captured by this simplified bracketing. Our treatment should therefore be interpreted as a first-order approximation rather than a definitive bound on IMF-driven variations.}

 {In general, the IMF variation is modelled via the mismatch parameter, defined as $\delta_{\rm IMF} = (M_{\rm \star}/L)_{\rm IMF}/(M_{\star}/L)_{\rm Chab}$, where $(M_{\star}/L)_{\rm IMF}$ and $(M_{\star}/L)_{\rm Chab}$ are the stellar mass-to-light ratios derived using a generic IMF and a Chabrier IMF, respectively. Our analysis relies on the results from~\cite{TRN13_SPIDER_IMF}, who applied Jeans dynamical modelling to a sample of local ETGs, assuming a reference model with an NFW dark matter halo profile~\citep{Navarro_1997}. From this work, we extract the median relation between $\delta_{\rm IMF}$ and $M_{\star, \rm Chab}$ as presented in~\cite{Tortora+2014}. Their findings indicate a mild dependence of the IMF on stellar mass  {($\delta_{\rm IMF} \approx 0.145\log (M_{\rm \star, Chab}/M_{\odot}) -0.305$)}.}

 {In the bottom-heavy IMF scenario, we expect an increase in the fraction of low-mass (dwarf) stars born within each galaxy. In this case, stellar feedback remains unchanged, and the structural properties, such as effective radii, are therefore not affected. The additional dwarf stars increase the stellar mass budget, but their low luminosity results in negligible changes to the galaxy light profile. Consequently, the size-luminosity ($R_{\rm e}-L$) relation is almost preserved relative to a Chabrier IMF, in agreement with~\cite{Barber_2018}. To implement this scenario, we modified the stellar mass according to:
\[
M_{\star, \rm IMF} = M_{\star, \rm Chab} \times \delta_{\rm IMF},
\]
while keeping luminosities and effective radii fixed. The updated stellar masses were then used to recompute velocity dispersions by adopting an approach similar to that used for computing $\sigma_{\rm e, dyn}$. Solving the Jeans equation with this modified two-component model, we calculate the velocity dispersion values under a bottom-heavy (BoH) IMF scenario, $\sigma_{\rm e, BoH }$.}

 {For the top-heavy (ToH) IMF scenario, the fraction of massive stars that are born is increased, which affects stellar feedback due to the higher number of short-lived massive stars and supernovae per galaxy. Stronger feedback can suppress star formation, leading to a decrease in present-day luminosity, while half-light radii may respond by increasing. However, the total stellar mass remains largely unchanged with respect to a \cite{Chabrier2003} IMF~\citep{Barber_2018, deGraaff+2022}. %particularly for massive galaxies (log $M_\star > 10.5$), as shown in ref.~\cite{Barber_2018}. 
To emulate this scenario, the luminosity was modified using the IMF mismatch factor ($\delta_{\rm IMF}$) such that:
\[
L_{\rm IMF} = \frac{L_{\rm Chab}}{\delta_{\rm IMF}},
\]
while the effective radii were increased by $0.1$ dex relative to its Chabrier-based counterpart (see \citealt{Barber_2018}). This modification of the luminosities and sizes %moves the galaxies ``to the left'' on the $L-R_{\rm e}$ plane, reproducing 
reproduces the expected $L-R_{\rm e}$ scaling behaviour of massive galaxies under a top-heavy IMF, i.e., fainter galaxies having larger effective radii, and hence much less surface brightness. This approach alters both the surface brightness profile and the aperture used to calculate the velocity dispersion, which, under this IMF scenario, we call $\sigma_{\rm e, ToH}$. These quantities are then updated consistently by applying the revised definitions introduced in this ToH framework.
%it does not affect the velocity dispersion of the galaxies, as the luminosity (or the corresponding surface brightness) profiles appear as weights in the Jeans equation (i.e., the mismatch factors are cancelled out, as they appear in both numerators and denominators).
}

 {We apply these different IMF scenarios to build the virtual-ETG FPs using Eq. (\ref{FP}), yielding four distinct FP parameterisations that account for the various IMF assumptions and velocity dispersion models. Below, we summarise the set of FP variables implemented in each of these models and the corresponding notation. The subscripts BoH and ToH denote parameters that are altered under the revised IMF prescriptions relative to the Chabrier reference model.
\begin{itemize}
    \item Chabrier IMF with $\sigma_{\rm corr}$: $(R_{\rm e},~\langle I_{\rm e}\rangle,~\sigma_{\rm e, corr})$
    \item Chabrier IMF with $\sigma_{\rm dyn}$: $(R_{\rm e},~\langle I_{\rm e}\rangle,~\sigma_{\rm e, dyn})$
    \item non-universal BoH IMF: $(R_{\rm e},~\langle I_{\rm e}\rangle,~\sigma_{\rm e, BoH})$
    \item non-universal ToH IMF: $(R_{\rm e, ToH},~\langle I_{\rm e}\rangle_{\rm ToH},~\sigma_{\rm e, ToH})$.
\end{itemize}}

%Finally, we used the photometric ($R_{\rm e}$ and $\langle I_{\rm e}\rangle$) and kinematic ($\sigma_{\rm e, corr},~\sigma_{\rm e, dyn}$ and $\sigma_{\rm e, \delta IMF}$) quantities introduced above to parameterize the virtual-ETG FPs according to Eq. (\ref{FP}).

% {Moreover, notice that this IMF variation is applied only in post-processing, by rescaling stellar masses according to empirical IMF–mass relations from dynamical studies of ETGs. As a result, it does not self-consistently modify stellar evolution, chemical enrichment, or feedback energetics in the simulation. The model therefore provides an observationally motivated test of the global kinematic impact of IMF variation on the FP, rather than a full feedback treatment. It also neglects possible radial IMF gradients within galaxies: observations show that massive ETGs often have more bottom-heavy IMFs toward their centres (e.g., ref. \cite{Bernardi+2023}), which would further increase central mass-to-light ratios and velocity dispersions. Incorporating such spatial variations self-consistently would require rerunning the simulation with a variable IMF, which is beyond the scope of this study.}

\subsection{Fitting Methods}
Following the different strategies in the literature (see, e.g.,~\citealt{Bernardi_2003b}), below we describe two of the most common fitting approaches to determine the free parameters $a,~b,~c$ of the FP. The first one is the direct fit, where one of the variables (in this case, the effective radius) is chosen as the dependent variable, and a linear regression is performed on the other two variables ($\sigma_{\rm e}$ and $\langle I_{\rm e}\rangle$). In other words, the objective of the direct fits is to minimise the cost function:
\begin{equation}
{\chi}_{\rm dir}^2 =  \sum_{i = 1}^{N}(\log R_{\mathrm{e},i} - a\log \sigma_{\mathrm{e}, i} - b\log\langle I_{\rm e}\rangle_i - c)^2.\label{eq:direct_fits}
\end{equation}
This method minimises the sum of squared residuals only in the direction of the chosen dependent variable.

The second fitting approach is the orthogonal fit, which minimises the sum of squared residuals in the direction orthogonal to the plane. This method finds the best-fitting plane that minimises the distance from all points to the plane in all three dimensions simultaneously, i.e., it treats all variables symmetrically.  In this case, the cost function to be minimised is:
\begin{equation}
 {\chi}_{\rm ortho}^2 = \sum_{i = 1}^{N}\frac{(\log R_{\mathrm{e},i} - a\log \sigma_{\mathrm{e}, i} - b\log\langle I_{\rm e}\rangle_i - c)^2}{{1+a^2+b^2}}.\label{eq:orthogonal_fits}
\end{equation}

%In evaluating these \S\ref{sec:Methods} for fitting the FP, it is important to highlight that our study currently lacks a comprehensive evaluation of the error covariance among the FP parameters. We suspect that these correlated errors might affect the results of orthogonal fits, but we have not confirmed the extent of this error correlation. Due to this uncertainty,
%, along with the reliability of our direct fits -- which minimize residuals in the $\log R_{\rm e}$ direction %and are crucial for using the FP as a robust distance indicator -- 
%our primary analysis was directed towards the direct fits. Specifically, with orthogonal fits, we get $b\sim -0.8$  consistent with literature, but get $a\sim1.8$, which is larger than the usually reported observational inferences, that vary more or less within the range $1.1 - 1.3$ (see, e.g., table 1 of ref.~\cite{Saulder+2013}). We recognize that a thorough investigation of the covariance matrix is essential to fully understand its impact on the FP fits, and we intend to undertake this in future research.

To find the FP parameters $a,~b,~c$ and related uncertainties, we used a least-squares algorithm combined with a bootstrapping approach (similar to the one used by~\citealt{Lu+2020}), which consists of iteratively resampling the data points with replacement and optimising the model parameters at each run. We perform the method described above with $500$ resampling iterations. Moreover, at each iteration, we clipped objects located at more than $3\sigma$ from the best-fit plane. This last procedure is commonly applied to data with a substantial impact of outliers (see~\citealt{Cappellari+2013} and references therein). Although the virtual-ETG catalogue is assumed to be mostly free of these extreme objects~\citep{ferreira2025cataloguevirtualearlytypegalaxies}, we chose to be extra careful and apply this criterion to ensure a better fitting stability. To quantify the concordance level between our results and the literature ones, we employ the (absolute value) Z-score:%the metric\cite{Bailey_2017}:
\begin{equation}
    Z = \frac{|x_1 - x_2|}{\sqrt{(\delta x_1)^2 + (\delta x_2)^2}},
\end{equation}
where $x_1 \pm \delta x_1$ and $x_2 \pm \delta x_2$ are two independent measurements of the same quantity, e.g., the coefficient $a$ or $b$ of the FP as performed by different works. A lower Z-score (typically less than 3) indicates a closer match, meaning the measurements differ by fewer standard deviations.

%For each one of these fitting \S\ref{sec:Methods}, we will compare our results with literature using the same method.

%\subsection*{External data.}
%To validate the simulated FP against observational data, we make use of several FP results derived in the literature, particularly, we have extracted the FP results in the $r$-band derived with direct fits from Saulder et al. (2013)\cite{Saulder+2013} and the summary of results provided in their table 1, which come from Bernardi et al. (2003c)\cite{Bernardi_2003} and Hyde \& Bernardi (2009)\cite{Hyde+2009}. Alongside this summary of results reported by Saulder et al., we additionally compare our findings with the more recent study by Bernardi et al. (2020)\cite{Bernardi+2020}.

\begin{table*} 
    \begin{center}
    \caption{\label{tab:fp_fits_label}  {Summary of the results obtained by fitting the virtual-ETG FP with different central velocity dispersion definitions (first column). We show the results of the fits for both direct (second column) and orthogonal (third column) fits. }} 
  \begin{tabular}{lcc}
    \hline
    \textbf{ {FP IMF model}} & \textbf{Direct (a, \,b, \,c)} & \textbf{Orthogonal (a, \,b, \,c)} \\
    \hline
     {Chab w/ }$\sigma_{\rm e, corr}$ & $(1.036 \pm 0.008, \; -0.561\pm0.003, \; 3.047 \pm 0.031)$  & $(1.330 \pm 0.011,\;-0.548\pm 0.003,\; 2.302\pm 0.038)$ \\
     {Chab w/} $\sigma_{\rm e, dyn}$  & $(1.176\pm 0.011, \; -0.709\pm 0.003, \; 3.879 \pm 0.030)$  & $(1.558\pm 0.012,\;-0.743\pm0.003,\; 3.301\pm 0.035)$  \\
     {BoH}  & $(1.107\pm 0.010,\;-0.726\pm 0.003, \; 4.133 \pm 0.026)$  & $(1.453\pm 0.012,\;-0.766\pm 0.003, \; 3.673 \pm 0.033)$  \\
     {ToH} & $(1.071 \pm 0.009,\; -0.685\pm 0.002,\;3.796 \pm 0.028)$ & $(1.362\pm0.011,\;-0.706\pm0.002 ,\; 3.302\pm0.032)$\\
    \hline
  \end{tabular}
\label{tab:fp_fits}
\end{center}
\end{table*}

\section{Results}
%\subsection*{Fundamental plane.} \label{sec:FP}
%\begin{figure*} 
%    \centering
%    \includegraphics[width=2\columnwidth]{figures/comparison_direct_fit_triple.pdf}
%    \caption{Edge-on view of the virtual-ETG fundamental plane from our sample from direct fits. In the left panel, the kinematic component of the FP is defined by $\sigma_{\rm e, corr}$, the middle panel displays the edge-on view of the Virtual-ETG FP defined with $\sigma_{\rm e, dyn}$, while the right panel presents the edge-on projection of the FP defined using $\sigma_{\rm e, \delta IMF}$. The grey dots are considered outliers (diverging more than $3\sigma$ from the best-fit plane) and thus are excluded from the fit. In both panels, we show the best-fit line (solid black) and the $1\sigma$ and $3\sigma$ deviations from the best-fit line (black dashed and dotted lines, respectively). %For each fitted plane, we also show the scatter (Root Mean Squared Error, RMSE) with uncertainties computed via boostrapping (see our \S\ref{sec:Methods} section). 
%    Lastly, the fitted points are coloured by their density, where redder (bluer) colours are regions with larger (smaller) density of objects.}
 %   \label{fp_edge_on_direct}
%\end{figure*}

%\begin{figure*}
%    \centering
%    \includegraphics[width=2\columnwidth]{figures/comparison_orthogonal_fit_triple.pdf}
%    \caption{Similar to Figure~\ref{fp_edge_on_direct}, but applying orthogonal fits to extract the FP parameters.}
 %   \label{fp_edge_on_orthogonal}
%\end{figure*}

In this section, we present the results of fitting the FP (\ref{FP}) to the virtual-ETG catalogue and compare the FP parameters with those reported in the literature. We focus our study on the photometric filter in the $r$-band. %To ensure a fair comparison, we have verified that all the observational samples used for reference provide $r$-band magnitude ranges that are compatible with those in our dataset.  

\subsection{Direct fits}
For the direct fit (Eq.~\ref{eq:direct_fits} in \S\ref{sec:Methods}), the optimal FP parameters employing the velocity dispersion definition $\sigma_{\rm e, corr}$ for the kinematic component, are identified as $a = 1.036 \pm 0.008$ and $b = -0.561 \pm 0.003$. In contrast, when the kinematic component of the FP is defined using $\sigma_{\rm e, dyn}$, the optimal coefficients are $a =1.176 \pm 0.011$ and $b = - 0.709 \pm 0.003$. 
%and $c = 4.046 \pm 0.036$, 
%The best-fit coefficients of the FP corresponding to these parameters is presented in the middle panel of Figure \ref{fp_edge_on_direct}. 
 {These results are summarised in  
%, and $c = 3.047 \pm 0.031$. 
%Figure 
Table \ref{tab:fp_fits}, 
%(first row, second column), 
where we show the best-fit FP coefficients obtained for the different velocity dispersion definitions and fitting methods}.  {For a visualisation of all results, we show the corresponding FP plots in Appendix A, Figure \ref{sup_fig_fp}.}
{Overall, the shift of parameters obtained using the $\sigma_{\rm e, dyn}$, especially the $b$ parameter, is significant, making the inferred FP nicely consistent with observations, as we will detail right below. This immediately suggests that the inclusion of the dynamically corrected central velocity dispersion, designed to mitigate the effect of the softening length, provides a convincing solution of the tension with observations, without invoking any modification of the feedback.}

{For instance, using the same photometric band for the IllustrisTNG100-1 simulation and applying similar galaxy classification criteria, \cite{Lu+2020} determined for ETGs at $z=0$, that the FP coefficients were $a= 1.295 \pm 0.026$ and $b= -0.627 \pm 0.012$. %and $c = 3.129 \pm 0.128$, exhibiting a mild evolution up to $z=0.1$. 
As discussed in the Lu et al. paper, these estimates did not match observations, which the author explained with the effect of the baryon physics on the assembly of the dynamical mass-to-light ration versus mass in simulations producing a wrong tilt. However, these results do not match} our findings for either the $a$ or $b$ coefficient, with both diverging by more than  $4\sigma$ from our predictions (whether using $\sigma_{\rm e, corr}$ or $\sigma_{\rm e, dyn}$ to characterize the FP's kinematic component, see also Table \ref{tab:sigma_agreement}, in Appendix B for a detailed comparison). 
%In theory, t
The main difference between the Lu et al. results and our findings resides in the FP observable definition: 
%s between the aforementioned FP parameters and our deductions could stem from two main factors: 1) the adoption of definitions in “observables” used for parameterising the FP, as Lu et al.\cite{Lu+2020} 
they use a projection of the stellar half-mass radius as a proxy for the photometric $R_{\rm e}$, a total luminosity in the $r$-band projected within this radius, and a softened $\sigma_{\rm e}$ (that is, defined without corrections for scales below the softening length).  {In fact, re-fitting our FP employing equivalent parameters and fitting procedure, we have found that $a = 1.229\pm 0.041$ and $b = -0.644 \pm 0.016$ at $z=0$, 
%for the galaxies projected along the $X$ axis (as in Lu et al.). These coefficients differ by less than $1.4\sigma$, 
i.e., largely consistent with the inferences of Lu et al., {thus demonstrating that using the proper definition of observables for the FP inference -- namely, implementing observational realism -- shifts the intrinsic FP encoded in  simulations quite substantially compared to an inference based on “blind” structural quantities derived from standard subfind catalogues.}} 

\begin{figure*} 
    \hspace{-0.3cm}
    \includegraphics[width=0.5\textwidth]{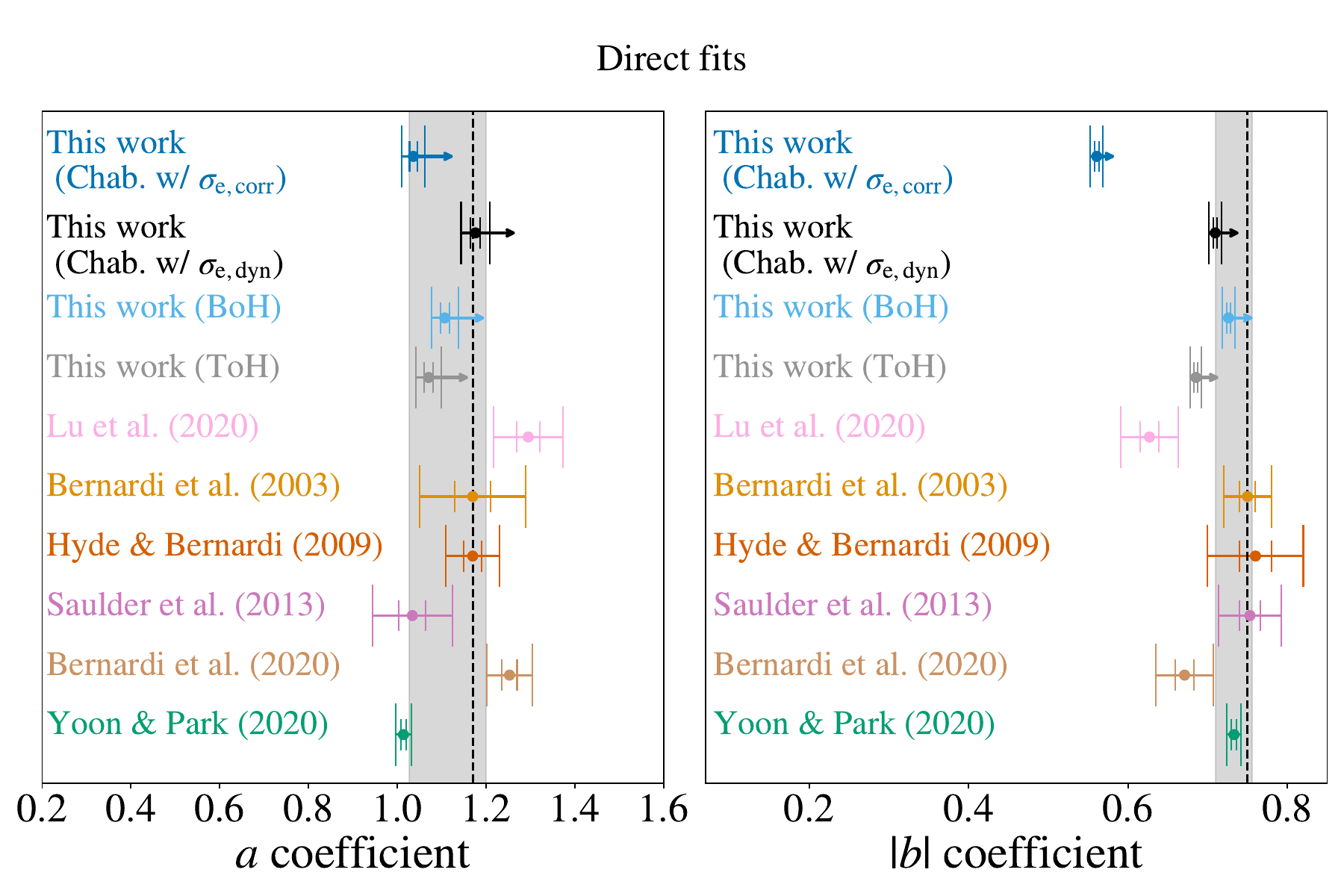}
     \includegraphics[width=0.5\textwidth]{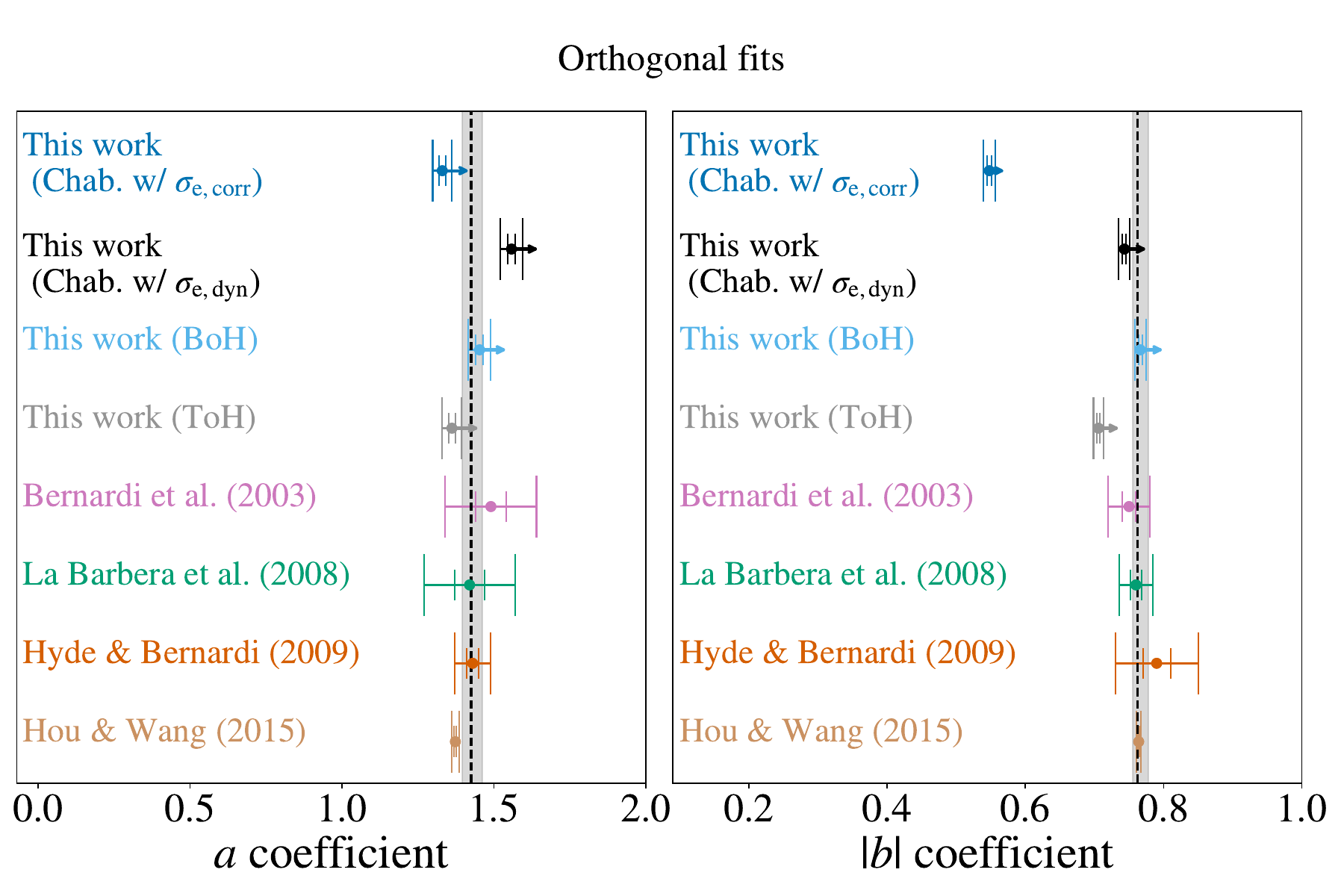}
    \caption{Best-fit coefficients of the fundamental plane from our analysis and literature for the {\it direct fit} (left) and the {\it ortogonal fit} (right). The small and larger error bars represent, respectively, the $1\sigma$ and $3\sigma$ confidence intervals of the inferences. The arrows in our inferences roughly represent how our results would change in the presence of ETGs having absolute $r$-band magnitudes fainter than $-20$.
    The vertical dashed line represents the median of the observational inferences, and the grey shaded region represents the 16th-84th percentiles.}
    \label{fig:summary_coefficients_FP}
\end{figure*}

Moving to the comparison of our inferences with observational ones in the $r$‑band, using a sample of $90\,000$ ETGs from the Sloan Digital Sky Survey (SDSS;~\citet{York+2000, Stoughton+2002}),\cite{Bernardi_2003b} found $a=1.17\pm0.04$ and $b=-0.75\pm0.01$ for $0.01<z<0.3$. Consistently, Hyde \& \cite{Hyde+2009} reported $a=1.170\pm0.054$ and $b=-0.757\pm0.022$ from $\sim50\,000$ ETGs in SDSS DR4~\citep{McCarthy-2006}. Using SDSS‑DR8~\citep{Aihara+2011}, \cite{Saulder+2013} obtained $a=1.034\pm0.030$ and $b=-0.753\pm0.013$ for $\sim93\,000$ ellipticals at $z<0.2$. Although slightly shallower, these agree with earlier values. More recently, \cite{Bernardi+2020} found $a=1.253\pm0.017$ and $b=-0.671\pm0.012$ for MaNGA ETGs~\citep{Bundy+2015}. This latter result has systematically larger (smaller) $a$ ($b$) coefficient than the previous observational results, which is likely due to their different FP parametrisation (using axis ratio as a third independent variable and half-light major axes of each ETG, instead of using the circularised effective radius). Lastly, \cite{Yoon_2020} determined, for $16\; 283$ SDSS-DR7~\citep{Abazajian_2009} ETGs at $0.025 \leq z <0.055$ that $a=1.01\pm0.006$ shallower than most observational results reported here and $b=-0.733 \pm 0.003$ consistent with other observations. Overall, all results show substantial agreement with the coefficient we found for the virtual-ETG sample. %as shown in Figure \ref{fig:summary_coefficients_FP}.
%summarises the findings: 
In particular, the derived FP parameters using $\sigma_{\rm dyn}$ show good agreement with observational data, within 1$\sigma$. In contrast, the best fit FPs with $\sigma_{\rm corr}$ consistently produce lower values of $|b|$ across all inferences and underestimate the $a$ parameter, aligning solely with results from \cite{Saulder+2013}, and, to a lesser extent, \cite{Bernardi_2003b} and \cite{Yoon_2020}.  {A comparison summary of the FP  coefficients from the Direct fits, using both the $\sigma_{\rm e, corr}$ and $\sigma_{\rm e, dyn}$, against the observed samples is shown in Figure \ref{fig:summary_coefficients_FP}.}

%\begin{figure*} 
%    \centering
%    \includegraphics[width=0.8\textwidth]{figures/updated_orthogonal_fits_summary_parameters.pdf}
%    \caption{Similar to Figure~\ref{fig:summary_coefficients_FP}, but for orthogonal fits.}
%    \label{fig:summary_coefficients_FP_orthogonal}
%\end{figure*}

\subsection{Orthogonal fits} For the orthogonal fits (Eq.~\ref{eq:orthogonal_fits}), the best-fit slopes of the FP employing $\sigma_{\rm e, corr}$ for the kinematic component are $a = 1.330 \pm 0.011$ and $b = - 0.548 \pm 0.003$, 
%On the other hand, 
while employing $\sigma_{\rm e, dyn}$, the best-fit FP slopes are $a = 1.558 \pm 0.012$ and $b= -0.743 \pm 0.003$. {As for the direct fit, the results obtained by the ``unsoftened'' $\sigma_{\rm e, dyn}$ return FP parameters fully aligned with observations, see below.} {All results of the FP orthogonal fit for the different velocity dispersion definitions  are also presented in Table~\ref{tab:fp_fits}.}
%(first and second row, third column).
Concerning the observational inferences, in the $r$-band, %Jorgensen et al. (1996)\cite{Jorgensen+1996} found $a = 1.24 \pm 0.07$ and $ b= -0.82 \pm 0.02$ for a sample of 226 ETGs in ten clusters. 
%Regarding the study conducted by 
%Instead, 
\cite{Bernardi_2003b}
%, specifically focusing on orthogonal fits, they 
%determined that the FP slope values were 
found $a= 1.490 \pm 0.05$ and $b = -0.75 \pm 0.010$.  \cite{LaBarbera+2008} have obtained $1.42 \pm 0.05$ and $b = -0.760 \pm 0.008$ for $1430$ ETGs by combining SDSS with UKIDSS data.
%and employing orthogonal fits. 
For a larger sample, composed of $46\,410$ SDSS ETGs, \cite{Hyde+2009} determined
%, via orthogonal fits, 
$a= 1.43 \pm 0.02 $ and $b = - 0.790 \pm 0.02 $. More recently
%, also for orthogonal fits, 
\cite{Hou_2015} have found $a= 1.373 \pm 0.004$ and $b = 0.764 \pm 0.001$ for $70793$ ETGs from SDSS DR7. The $a$ coefficient obtained by these observational results is not in stark disagreement with the virtual-ETG FP derived using $\sigma_{\rm e, corr}$, but significantly diverges for the $b$ coefficient, as in the case of direct fits. On the other hand, the observational FPs are in broad agreement with the ones obtained for the virtual-ETG FP defined using $\sigma_{\rm e, dyn}$ for both coefficients, although the $a$ coefficient is systematically higher than the observational results.  {This is visualised in Figure \ref{fig:summary_coefficients_FP} (pair of plots on the right), where we compare the FP  coefficients from the Orthogonal fits against the observed samples, as above.}

\subsection{Impact of the luminosity range}
{The virtual-ETG sample is brighter than most of the ones from the literature used above, being the $r$-band absolute magnitudes of the former $M_{\rm r} \lesssim -20$, while the majority of the observational samples have $M_{\rm r} \lesssim -19$. However, following \cite{Hyde+2009}, who have estimated how the FP parameters are affected as a function of the depth of the faint end, we estimate that the $a$ parameter increases by $\sim0.1$ while the $b$ parameter increases by $\sim5$ per cent in both fitting approaches. These offsets are represented by the arrow tips in Figure~\ref{fig:summary_coefficients_FP} %and \ref{fig:summary_coefficients_FP_orthogonal}, 
%in which the results of our comparisons with the literature are presented. We emphasise that these modifications do not alter the qualitative 
showing that our results are not significantly affected by the use of a brighter sample.}  {For a more quantitative assessment, since our sample does not contain galaxies fainter than $M_{\rm r} \simeq -20$, we performed the complementary test of applying the same luminosity cut to an observational catalogue. Specifically, we re-fitted the FP using the DynPop ETG sample~\citep{Zhu} restricted to $M_{\rm r} \leq -20$. The resulting FP coefficients become slightly shallower -- particularly the $a$ coefficient -- which is consistent with our expectations for luminosity-limited samples. However, the magnitude of these shifts is comparable to, or smaller than, the statistical uncertainties of the best-fit FP coefficients, indicating that luminosity range does not alter our main conclusions.
This is shown in the Appendix (Figure \ref{fig:extended_data_fig_direct}). 
%Figures~\ref{fig:extended_data_fig_direct} and \ref{fig:extended_data_fig_orthogonal}. 
%The magnitude of these shifts is comparable to, or smaller than, the statistical uncertainties of the best-fit FP coefficients, indicating that luminosity range does not alter our main conclusions.
}

\section{Discussion}
As seen earlier, 
%The findings regarding the FP inferences presented previously, including both simulations and observations, are summarized in 
Figure~\ref{fig:summary_coefficients_FP}
%and \ref{fig:summary_coefficients_FP_orthogonal} 
shows a comparison of the FP coefficient inferences with observations and independent IllustrisTNG analyses, for both Direct and Orthogonal fits, respectively. 
 {Looking at these results at face value, the simulated FP exhibits slopes that differ from the observed FP by $\Delta a\sim 0.05$ and $\Delta b\sim 0.1$. Although these differences may appear small, for typical ETG values ($\log \sigma_{\rm e}\sim2.4$, $\log I_{\rm e}\sim2.5$), even such modest changes imply $\Delta \log(M_{\rm dyn}/L)\sim0.25$ dex, i.e., up to 75 per cent change in inferred dynamical mass-to-light ratios. Thus, even ``small'' FP slope differences correspond to large systematic shifts in dynamical mass inferences. %(of the order of the IMF changes). 
This reinforces the extreme importance of proving that simulated FPs are as close to observations as possible.}
 {This is even more important if we look at the statistical deviance of the simulated FP inferences from the median observed ones. This is quantified in} 
%For a more quantitative comparison, in 
Table~\ref{tab:median_z_scores}  {where} we present the median agreement level, quantified via the median Z-score, i.e. how far our parameter inferences are from the different literature results reported in the previous section, taking into account measurement errors (see \S\ref{sec:Methods}). 
%between our inferences and the others from the literature. 
For the individual Z-score results, we refer the reader to Table \ref{tab:sigma_agreement}. 
Our findings show that using $\sigma_{\rm e, corr}$ for the FP kinematic parametrisation yields $a$ and $b$ coefficients strongly deviating from
%that do not generally meet 
observational inferences. %, aside from the $a$ coefficients by Saulder et al. (2013)\cite{Saulder+2013} for direct fits and Jorgenssen et al. (1996)\cite{Jorgensen+1996} for orthogonal fits. 
The limitation of $\sigma_{\rm e, corr}$ is even more evident for the $b$ coefficient, which deviates from observational medians by over $14\sigma$ and $22 \sigma$ for the direct and orthogonal fit, respectively (see Table \ref{tab:median_z_scores}).  %failing to meet observational expectations in all fitting \S\ref{sec:Methods} (Figures~\ref{fig:summary_coefficients_FP} and~\ref{fig:summary_coefficients_FP_orthogonal}). 
This suggests our weighted velocity dispersion modification, $\sigma_{\rm e, corr}$, is still too weak for smaller galaxies for which the potential softening can impact radii outside the softening length, which we have assumed to have correct kinematics when calculating $\sigma_{\rm e, corr}$.  {This dispersion depletion remains despite our $\sigma_{\rm e, corr}$ model being designed to maximise the correction inside the effective radius (see \S\ref{sec:Methods})}. 
% {Moreover, the inability of this corrected dispersion to recover the observed FP is also plausibly related to its neglect of structural variations among galaxies, in particular the dependence on Sérsic index (see \S\ref{sec:Methods})}. 
On the other hand, using the dynamically inferred central velocity dispersion $\sigma_{\rm e, dyn}$, the FP aligns much better with observations for any fitting method, as shown in Figure~\ref{fig:summary_coefficients_FP} %\ref{fig:summary_coefficients_FP_orthogonal}
 and Table~\ref{tab:median_z_scores}. %The direct fits reveal a median deviation of $\sim2.6\sigma$ for the $a$ coefficient and $\sim2.2\sigma$ for the $b$ coefficient. Notably, the deviations in orthogonal fits are even smaller, with median Z-scores of $\sim1\sigma$ for the $a$ coefficient and $\sim1.2\sigma$ for the $b$ coefficient.

%Remarkably, % by comparing our derived inferences against observational data, 
%we observe a broad agreement when using the dynamically inferred central velocity dispersion $\sigma_{\rm e, dyn}$ for describing the FP's kinematic component. It is evident from Table~\ref{tab:sigma_agreement} that this definition is consistent with the observational FPs,  showing median deviations of $\sim 2.4\sigma$ for the $a$ coefficient and $\sim3.1\sigma$ for the $b$ coefficient. This holds true for all inferences, except for Saulder et al. (2013)\cite{Saulder+2013} in the $a$  coefficient, which deviates by slightly more than $4\sigma$, and for Bernardi et al. (2020)\cite{Bernardi+2020}, that used a different FP parameterization, making direct comparison less meaningful.

%\section*{Conclusions}
%\label{sec:FP_discrepancy}
\subsection{Resolution effects} The takeaway message is that the IllustrisTNG100-1 
simulation {, and hydro-simulations in general, require} a velocity dispersion definition that minimises sub-softening scale effects to replicate the ETG FP in real galaxies. Higher spatial resolution simulations (with softening lengths of about a tenth of a kpc)  {and sufficiently large cosmological volumes are expected to better reproduce observations by reducing sub-softening effects, and by more accurately sampling the high-mass end of the galaxy population, which is otherwise under-represented in smaller volumes (e.g., the TNG50-1 simulation~\citealt{Nelson+2019, Pillepich+2019}).}
%can improve observational matches without assigning dynamically inferred velocity dispersion to sub-halos 
 {Consistent with this, our resolution-convergence test in the Appendix shows that TNG100-1 and TNG50-1 yield nearly identical kinematic and density profiles at $r \gtrsim 1$ kpc, but diverge at shorter radii, precisely where our dynamical modification kicks in to reproduce the ``unsoftened'' velocity dispersion profile.
%However, it is important to mention that although higher-resolution simulations such as TNG50-1\cite{Pillepich+2019, Nelson+2019} offer higher spatial resolution than TNG100-1, its $\sim 10 \times$ smaller volume undersamples the massive ETG population that dominates observational FP studies. Because FP slopes depend systematically on galaxy mass, TNG50-1 does not provide a statistically representative comparison. Furthermore, our resolution converergence test (Supplementary Figure \ref{fig:resolution_convergence_comparison}) shows that TNG100-1 and TNG50-1 yield consistent kinematic and density profiles at $r \gtrsim 1$ kpc, the scales relevant for our FP fits. For this reason, we base our FP analysis on TNG100-1 while acknowledging that complementary high-resolution studies are a promising direction for future work.
}

\begin{table} 
    \begin{center}
    \caption{\label{tab:l1}Median Z-scores between simulated and observed FPs. Lower values indicate better agreement. We present the results derived for the FP characterised by the several parameterisations we introduced in the main text (first column) and for the two fitting methods, i.e., direct (second column) and orthogonal (third column). Values shown in bold indicate the strongest alignment with the observed FP.} 
  \begin{tabular}{lcc}
    \hline
    { {FP IMF model}} & {Direct (a,\,b)} & {Orthogonal (a,\,b)} \\
    \hline
    {Chab w/} $\sigma_{\rm corr}$ & $(2.527, \; 14.486)$  & $(3.472,\;22.251)$ \\
    {Chab w/} $\sigma_{\rm dyn}$  & $(3.817, \; 3.287)$  & $(4.058,\;2.146)$  \\
    \textbf{BoH}  & $\mathbf{(2.306,\;2.022)}$  & $\mathbf{(0.849,\;0.979)}$  \\
    {ToH} & $(2.420,\; 5.129)$ &  $(1.822,\;5.403)$ \\
    \hline
  \end{tabular}
\label{tab:median_z_scores}
\begin{flushleft}
Note: As this table aims to compare our results with observations, we omit the values obtained by \cite{Lu+2020} for the TNG100-1 FP.
\end{flushleft}
\end{center}
\end{table}

This velocity dispersion, alongside S\'ersic parameters $R_{\rm e}$ and $\langle I_{\rm e}\rangle$, is pivotal in defining the virtual-ETG FP and here we showed, for the first time, that using these realistic observable quantities for the simulated ETGs resolves discrepancies with observational FPs. %by employing the correct definitions of the simulated “observables”. 
%This is evident from Figures \ref{fig:summary_coefficients_FP}, \ref{fig:summary_coefficients_FP_orthogonal} and the Extended Data Table 1. %In summary, 
%This was already found in previous analyses (e.g., \cite{Lu+2020,2021_horizon,deGraaff+2022}), however, 
%the novelty of our analysis lies in minimizing the impact of how the observables involved in Eq. (\ref{FP}) are defined (e.g., 2D vs. 3D quantities) and measured (e.g., S\'ersic fitting) as well as accounting for the effects of numerical resolution (e.g., adaptive partitioning, $\sigma_{\rm e}$-correction). This robust methodology allowed us to align the FP inferences from the virtual-ETG catalogue with observed samples. 

%\noindent

\subsection{Virial consistency and feedback}\label{sec:virial_consistency_and_feedback}

The agreement of the corrected FP coefficients with observations does not automatically guarantee that the underlying galaxy structure is physically realistic. A further and independent test is therefore to verify whether the recovered FP coefficients are dynamically consistent with the other structural and kinematical scaling relations of the same virtual-ETG sample. In particular, we investigate whether the observed FP tilt can be reproduced within a weak-homology regime, or whether it would instead require an unrealistically strong luminosity dependence of the effective virial coefficient.

Starting from the scalar virial theorem,
\begin{equation}
M_{\rm dyn} \propto K_V \sigma_{\rm e}^2 R_{\rm e},
\end{equation}
and assuming $M/L \propto L^\alpha$, together with $L \propto I_{\rm e} R_{\rm e}^2$, the FP relation
\begin{equation}
R_{\rm e} \propto \sigma_{\rm e}^a I_{\rm e}^b
\label{eq: FP_th}
\end{equation}
can be derived under the assumption of strict homology ($K_V={\rm const}$), yielding
\begin{equation}
a_{\rm hom} = \frac{2}{1+2\alpha},
~~~~b_{\rm hom} = -\frac{1+\alpha}{1+2\alpha}.
\end{equation}

More generally, if the effective virial coefficient varies systematically with luminosity according to
$K_V \propto L^{-\kappa}$, the effective FP tilt becomes controlled by the combined parameter $\delta = \alpha + \kappa$, such that
\begin{equation}
a = \frac{2}{1+2\delta},
~~~~
b = -\frac{1+\delta}{1+2\delta}.
\end{equation}
Using the slope of the measured $M/L-L$ relation of the virtual-ETG sample ($\alpha\sim0.2$ see Appendix \ref{appendix:2d_scaling_relations}) and assuming strict homology ($\kappa=0$), we obtain predicted FP coefficients broadly comparable to the orthogonal-fit determinations, but still systematically offset. For the typical slopes measured in the virtual ETGs, the homology expectation gives approximately $a_{\rm hom} \sim 1.3-1.4$,
$|b_{\rm hom}| \sim 0.82-0.86$, while the orthogonal FP fits yield values around $a_{\rm orth} \sim 1.45-1.56$, 
$|b_{\rm orth}| \sim 0.74-0.77$.

The virial expectation therefore reproduces reasonably well the observed range of the $a$ coefficient, while generally predicting somewhat steeper values of $|b|$ than inferred from the FP fits. The residual discrepancy is particularly important because it indicates that the FP tilt cannot be fully described through a purely luminosity-dependent $M/L$ term alone.
The remaining tilt can be interpreted as an effective luminosity dependence of the virial coefficient. Introducing the empirical luminosity scalings $\sigma_{\rm e} \propto L^s$, and $R_{\rm e} \propto L^\eta$ in Eq. (\ref{eq: FP_th}) the virial closure becomes
\begin{equation}
\kappa_{\rm req} = \eta + 2s - 1 - \alpha.
\end{equation}
Using again the scaling relations measured for the corrected virtual-ETG sample (see Appendix \ref{appendix:2d_scaling_relations}), we estimate typical values of $s \sim 0.23$ and $\eta \sim 0.75$, thus leaving $\kappa_{\rm req}$ to remain small, typically of order $\kappa_{\rm req}\sim 0-0.1$.
This implies an effective tilt parameter
$\delta=\alpha+\kappa_{\rm req}$, which, for the measured $M/L-L$ slopes, predicts FP coefficients of order $a_{\rm pred}\sim1.3-1.4$, $|b_{\rm pred}|\sim0.8-0.85$.

Thus, the virial closure reproduces a substantial fraction of the observed FP tilt, and in particular gives values of $a$ comparable to the orthogonal FP determinations. In principle, this agreement could still arise from a degeneracy among the scaling-relation slopes conspiring to produce an approximately correct FP closure. However, we tend to disfavour this interpretation because the individual scaling relations themselves are also broadly consistent with observations, as discussed in \citet{ferreira2025cataloguevirtualearlytypegalaxies}. In particular, the virtual-ETG sample successfully reproduces several independent observables, including surface photometry, total density slopes, and galaxy kinematics, providing a level of global consistency not previously achieved simultaneously by other simulation catalogues.

However, taking the predicted coefficients at face value, we find that the expected $|b|$ values remain slightly larger than the fitted values, $|b|\simeq0.74-0.77$. This residual difference indicates that a one-parameter luminosity-dependent description of $M/L$ and $K_V$ is not sufficient to capture the full FP geometry. Additional covariance with compactness, surface brightness, total density slope, dark matter fraction within $R_{\rm e}$, or orbital structure is likely required and could be tracked to missing physics on the feedback.

Nevertheless, the weak values of $\kappa_{\rm req}$ inferred from the virial closure ($\kappa_{\rm req}\sim 0-0.1$) imply that only mild departures from homology are needed. In this sense, the present analysis suggests that the current TNG feedback implementation is already broadly compatible with the dynamical structure of observed ETGs \citep[e.g.,][]{Cappellari+2006,Auger+2010,Cappellari+2013}, and that the FP does not require major revisions of the feedback model. Any substantial modification of stellar, SN, or AGN feedback would need to preserve simultaneously the full set of scaling relations of the virtual-ETG sample (including the FP, surface photometry, density slopes, and kinematics) which considerably reduces the available freedom for large feedback changes. Rather, the remaining discrepancies are more likely associated with second-order effects regulating the detailed redistribution of baryons and dark matter within the effective radius, or with residual observational-realism uncertainties affecting the mapping between simulated and observed galaxy properties.

\subsection{Non universal IMF} However, feedback is not the only factor that can affect the FP inferences. As mentioned in the Introduction, another ingredient of stellar physics, which is usually ignored
%that could significantly affect the FP inferences is 
%overlooked 
in standard hydrodynamical simulations, is the non-universality of the IMF.  {This has been only touched in rare post-processing analyses of hydrodynamical simulations (e.g., \citealt{Barber_2018, deGraaff+2022}).} Unlike simulations, including IllustrisTNG100-1, real galaxies show a variation of the IMF, which is a function of the stellar mass or velocity dispersion, thus adding an extra term in the $M_{\rm dyn}/L-M_{\rm dyn}$ relation, which will contribute to the tilt in the FP~\citep{Binney}.  {This is usually parametrised by the so-called mismatch parameter $\delta_{\rm IMF}$, which is defined by the stellar mass-to-light ration deviation from a Chabrier IMF, such that steeper IMF at the low-mass end, like a \cite{Salpeter+1955} IMF, have a $\delta_{\rm IMF}>1$ (see \S\ref{sec:Methods}).}

 {We can try to emulate the impact of the IMF variation on the FP by forward modelling the effect of the IMF on the structural and dynamical properties of galaxies. Indeed, because of the change in the relative fraction of low-mass and high-mass stars, a variable IMF can cause a variation on the overall feedback models, e.g., by changing the number and power of SN, stellar winds and overall AGN feedback per unit of $M_\star$, with respect to a universal IMF~\citep{Gutcke+2018,Ziegler+2022}. Post-processed hydro-simulations~\citep{Barber_2018} have shown that a non-universal IMF depending on gas pressure, density and temperature, in either a top-heavy (ToH) or a bottom-heavy (BoH) IMF variation over a standard Chabrier IMF, could produce a realistic $\delta_{\rm IMF}-\sigma$ relation, compatible with observations~\citep{Cappellari+2013b}.  
In our toy-model, we try to cover these two extreme scenarios. In particular, in \S\ref{sec:Methods}, we discuss that, under the ToH scenario, galaxies tend to decrease their luminosities and increase their projected sizes, while there is not much impact on their stellar mass. Hence, the FP can experience a tilt due to the variation of the $R_{\rm e}$ and $I_{\rm e}$, which also induces a change on the $\sigma_{\rm e}$ due to the different aperture. In the BoH scenario, on the other hand, the increased fraction of the low-mass stars directly impact the stellar mass of the galaxies, which we quantify using a $\delta_{\rm IMF}$ and $M_{\rm \star, Chab}$ relation found in observations (in particular, the one from \citealt{Tortora+2014}), i.e., $\delta_{\rm IMF} \approx 0.145\log (M_{\rm \star, Chab}/ M_{\odot}) - 0.305$. This is used to renormalise the stellar density and luminosity profiles before solving the Jeans equation for the corresponding $\sigma_{\rm e, \delta_{\rm IMF}}$. To distinguish the modified $\sigma_{\rm e, \delta_{\rm IMF}}$ in the two scenarios we will indicate them as $\sigma_{\rm e, {\rm ToH}}$, and $\sigma_{\rm e,{\rm BoH}}$, respectively.}

 {Including these new IMF emulations on our FP parameters, we have the following FP slope values for the BoH scenario: $a=1.107\pm0.010$ and $b=-0.726\pm0.003$ for direct fits and $a=1.453\pm0.012$, $b=-0.766\pm0.003$ for orthogonal fits. On the other hand, for the ToH scenario we find $a = 1.071\pm 0.009$ and $b=-0.685 \pm 0.002$ for direct fits and $a=1.362\pm0.011$, and $b = -0.706\pm0.002$ for orthogonal fits.}
%Including the IMF‑adjusted dispersion $\sigma_{\rm e,\delta\rm IMF}$ in the virtual‑ETG FP yields . 
 {These results are summarised in Table~\ref{tab:fp_fits} (third and fourth rows)} and a full coefficient comparison is also reported in Figure \ref{fig:summary_coefficients_FP}. 
 %and \ref{fig:summary_coefficients_FP_orthogonal}.  
 {It is important to mention that, when adopting non-universal IMF assumptions, the directions of the coefficient changes relative to the Chabrier baseline agree with those found in the emulation of non-universal IMFs for the EAGLE model (see \citealt{Barber_2018} and fig. 17 from \citealt{deGraaff+2022}).}

 {For the BoH scenario, the median deviations from observational inferences (reported in Table~\ref{tab:median_z_scores}) are $2.31\sigma$ and $2.02\sigma$ for $a, b$ derived through direct fits, and $0.85\sigma$ and $0.98\sigma$ for $a, b$ derived via orthogonal fits. For the ToH scenario, the median deviations are $2.42\sigma$ and $5.13\sigma$ for $a,b$ derived through direct fits and $1.82\sigma$ and $5.40\sigma$ for $a,b$ derived through orthogonal fits. 
Taking these results in a relative scale to the case of the $\sigma_{\rm e, dyn}$ under the ``universal'' Chabrier IMF,
%In particular, under the ToH IMF scenario, 
the tension in the $b$ parameter is increased by about $3\sigma$ for the ToH model,
%relative to the Chabrier IMF with $\sigma_{\rm e, dyn}$ case, 
whereas under the BoH IMF scenario, the agreement with the observed FPs has improved significantly.}

{Interestingly, this improvement also appears qualitatively consistent with the virial-consistency analysis discussed in \S\ref{sec:virial_consistency_and_feedback}. Using the scaling relations measured for the BoH case (reported in Appendix~\ref{appendix:2d_scaling_relations}), we find a steeper $M/L-L$ relation, corresponding to $\alpha \sim 0.25-0.30$, while the luminosity dependences of the velocity dispersion and effective radius remain approximately unchanged, with $s \sim 0.23$ for the $\sigma_{\rm e}-L$ relation and $\eta \sim 0.75$ for the $R_{\rm e}-L$ relation.
These values imply a virial closure requiring only a very weak luminosity dependence of the effective virial coefficient, corresponding to $\kappa_{\rm req} \sim -0.05$ to $0.0$. The resulting effective tilt parameter therefore remains around $\delta \sim 0.25$, yielding expected FP coefficients of approximately $a_{\rm pred} \sim 1.3$ and $|b_{\rm pred}| \sim 0.8$.

Although these latter values do not reproduce perfectly the observed orthogonal-fit coefficients, they move significantly closer to the observed FP regime while still remaining compatible with the weak/no-homology interpretation inferred from the other structural and dynamical scaling relations. In this sense, the BoH case provides a better compromise between the observed FP tilt and the requirement that the effective virial coefficient varies only weakly along the galaxy sequence, reducing the need for strong structural non-homology.}

% {For comparison, using the Chabrier-based IMF $\sigma_{\rm e,\rm dyn}$ produces a $b$ that is $3.29\sigma$ too shallow in direct fits and an $a$ that exceeds observational medians by $4.06\sigma$ in orthogonal fits.  This demonstrates that non-universal IMFs, in particular, under a bottom-heavy scenario, %$\sigma_{\rm e,\delta\rm IMF}$ 
%greatly improves orthogonal fits, while only minimally affecting the direct‑fit $a$ coefficient, and thus 
%can provide a more realistic FP regardless of the fitting method.}

%{\it What are the possible sources of bias regarding our FP results?} 

To conclude, this control test demonstrates that introducing some “observationally motivated” IMF variations into the FP parameters can produce a further ``tilt'' in the simulated FP.
%,  {that in the case of the BoH model is} fully consistent with observations. 
 {Interestingly, our analysis has also shown that, if on one hand, the ToH variation moves the FP coefficient farther from the observations, than the standard Chabrier IMF, on the other hand, for the BoH IMF variation, the forward modeling shifts the velocity dispersion distribution in a way to produce a FP fully consistent with observations, without altering the current Illustris TNG feedback.}
%, producing the closest-to-observations FP, 
% {reinforcing our argument} 
%evidence 
%that feedback  {(albeit incomplete)}
%changes 
%does not primarily drive the FP discrepancy. }
% {As discussed earlier, this does not imply that the IllutrisTNG100-1 feedback model is correct, it rather suggests that, once sufficient observational realism is introduced into the simulation outputs, 
%we can more safely identify 
%the remaining discrepancy can more safely tracked to shortcomings of the baryonic model.}

 %and thereby it 

%\noindent
\section{Conclusions}  

{Taken together, our results suggest that the discrepancy between simulated and observed FP relations may not arise primarily from shortcomings in baryonic physics, such as feedback prescriptions or baryon-dark matter ratios \citep{Y.Wang, Lu+2020, deGraaff+2022}, but instead is strongly influenced by the way galaxy observables are defined and measured.

While sample selection and FP parameterisation introduce secondary effects, we find that incorporating observationally motivated definitions of galaxy properties -- particularly for the velocity dispersion -- substantially reduces the tension between simulations and observations. This highlights the importance of kinematic realism when interpreting scaling relations in hydrodynamical simulations.

In addition, we find that deviations from strict structural homology and systematic trends in the stellar mass-to-light ratio along the FP both contribute non-negligibly to the observed tilt. Variations in S\'ersic index and dynamical structure with mass or luminosity can account for part of the tilt, but our analysis shows that plausible M/L gradients driven by stellar population variations are required to reproduce the full amplitude and mass dependence of the FP tilt. The combined effect of non-homology and M/L trends therefore plays a central role in reconciling simulated and observed planes, and must be modelled jointly with the kinematic measurement choices.

Within this framework, variations of the stellar IMF introduce an additional degree of freedom that is largely degenerate with standard feedback prescriptions and with the inferred M/L trends. In particular, we show that a bottom-heavy IMF scenario can bring the simulated FP into close agreement with observations. However, this result should be interpreted as a demonstration of degeneracy rather than a unique solution, as similar effects may arise from alternative modifications to the underlying physical models, including different implementations of structural evolution and stellar population gradients.

Looking ahead, improvements in numerical resolution and dynamical modelling will further enhance the fidelity of simulated galaxy properties, and will allow a more direct treatment of non-homology and spatially resolved M/L variations. At the same time, incorporating IMF variability in a self-consistent manner remains an important open challenge for galaxy formation models. Addressing this aspect will be essential for disentangling the relative roles of baryonic physics, stellar populations, structural non-homology, and measurement systematics, and for developing predictive models capable of reproducing the full set of observed scaling relations of ETGs.}

\section*{Acknowledgements}
PdAF gratefully acknowledges the financial support provided by the Coordination of Superior Level Staff Improvement -- CAPES, grant No. 88887.140224/2025-00.
%NRN acknowledges support from the Guangdong Science Foundation grant (ID: 2022A1515012251). 
The authors gratefully acknowledge Luigi Giaquinto for his helpful discussions and assistance.

\section*{Data Availability}
The virtual-ETG catalogue used in this work is described and validated in \cite{ferreira2025cataloguevirtualearlytypegalaxies}, and is publicly available at \url{https://github.com/PedroFerreirAstro/TNGLDP_catalogue_ETGs}.
%\section*{Code Availability}
The data analysis in this work was performed with several Python libraries as NumPy, Matplotlib, Seaborn and SciPy. These packages are publicly available through the Python Package Index ({\url{https://pypi.org}}).
Specifically, the fit analysis in this study is performed using the Python package scipy.least\_squares. %We provide the .ipynb file for reproducing the images presented in the paper at the GitHub url: \url{https://github.com/PedroFerreirAstro/Virtual_ETG_Fundamental_Plane}.

\section*{Contributions}
PdAF contributed to the sample preparation, modelling, data analysis, and writing of the manuscript; NRN contributed to the sample preparation, data analysis, interpretation, and revision of the text; CT contributed to the modelling of data and revision of the text; LC and FVN contributed to the interpretation of results and revision of the text.

%\section*{Competing interests}
%The authors declare no competing interests.



\bibliographystyle{mnras}
\bibliography{example} % if your bibtex file is called example.bib

\appendix
\section{Supplementary Figures}

\begin{figure*}
    \includegraphics[width=\textwidth]{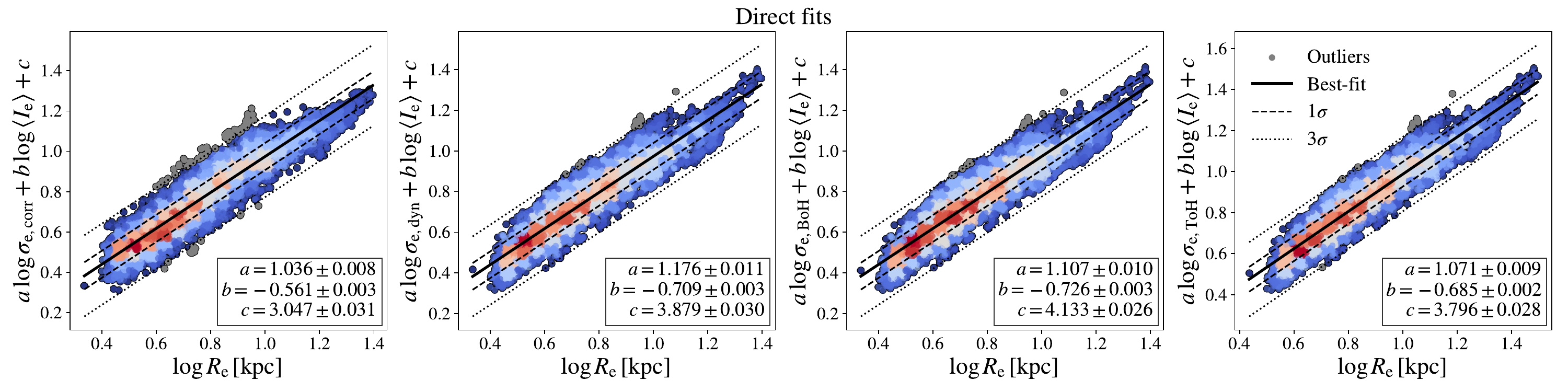}
    \includegraphics[width=\textwidth]{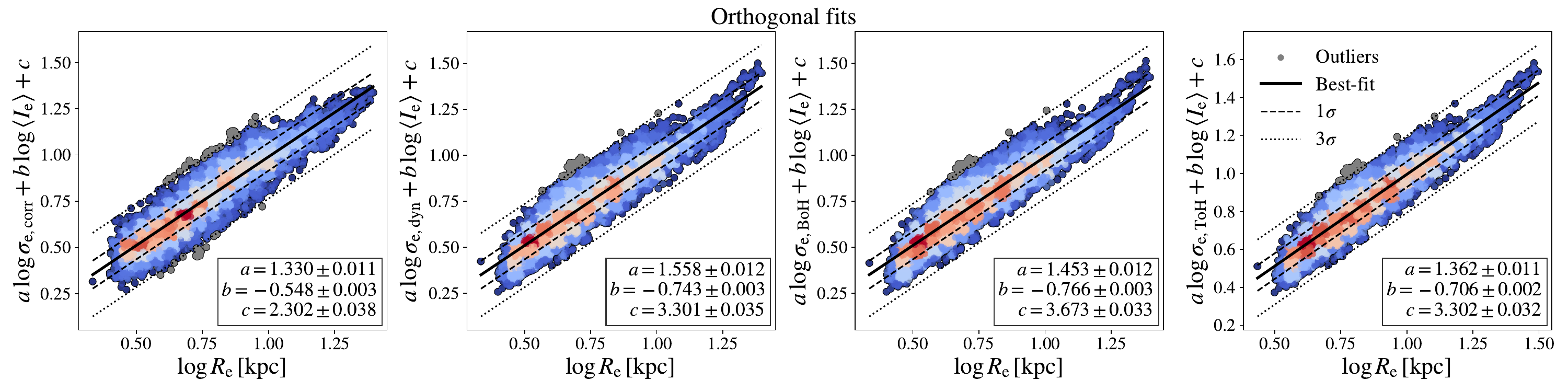}
    \caption{ {Edge-on view of the virtual-ETG fundamental plane from our sample from direct fits (upper row) and orthogonal fits (bottom row). In the first column (left to right), the kinematic component of the FP is defined by the Chabrier-based parameters with $\sigma_{\rm e, corr}$, while the second one is the Chabrier-based FP with $\sigma_{\rm e, dyn}$. The last two columns correspond to the FPs defined under non-universal IMF assumptions  (BoH and ToH, respectively) by using our forward modelling. The grey dots are considered outliers (diverging more than $3\sigma$ from the best-fit plane) and thus are excluded from the fit. In both panels, we show the best-fit line (solid black) and the $1\sigma$ and $3\sigma$ deviations from the best-fit line (black dashed and dotted lines, respectively). %For each fitted plane, we also show the scatter (Root Mean Squared Error, RMSE) with uncertainties computed via boostrapping (see our \S\ref{sec:Methods} section). 
    Lastly, the fitted points are coloured by their density, where redder (bluer) colours are regions with larger (smaller) density of objects.}}
    \label{sup_fig_fp}
\end{figure*}

 {\subsection*{Edge-on projection of the virtual-ETG FP}
In Figure \ref{sup_fig_fp} we show the edge-on projections of the best-fit FP to each of the four models we tried, i.e., Chabrier IMF with $\sigma_{\rm e, corr}$ and $\sigma_{\rm e, dyn}$, and the non-universal IMF emulations, BoH and ToH.}

\begin{figure*}
% \centering
\hspace{-0.4cm} 
\includegraphics[width=0.5\textwidth]{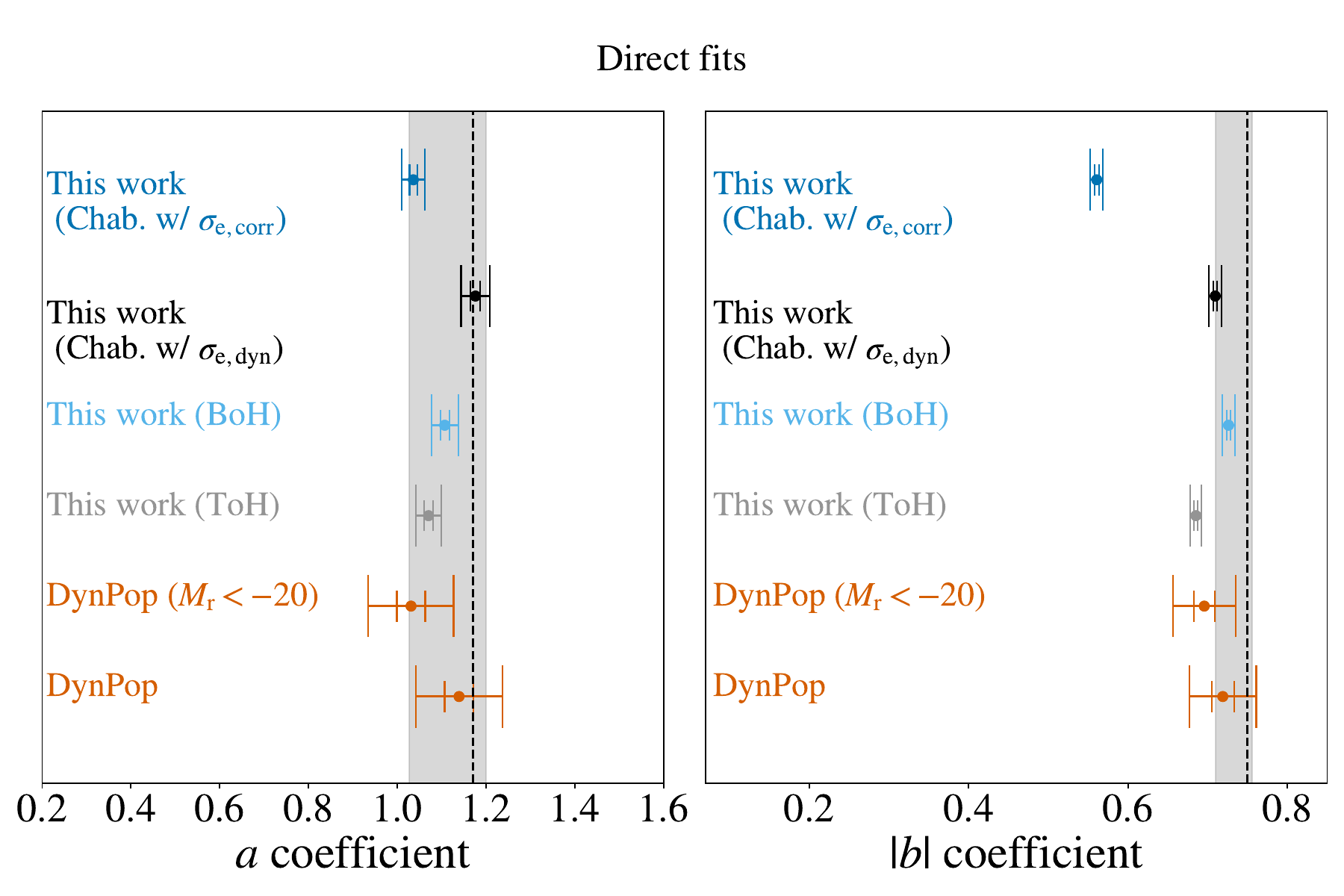}
\includegraphics[width=0.5\textwidth]{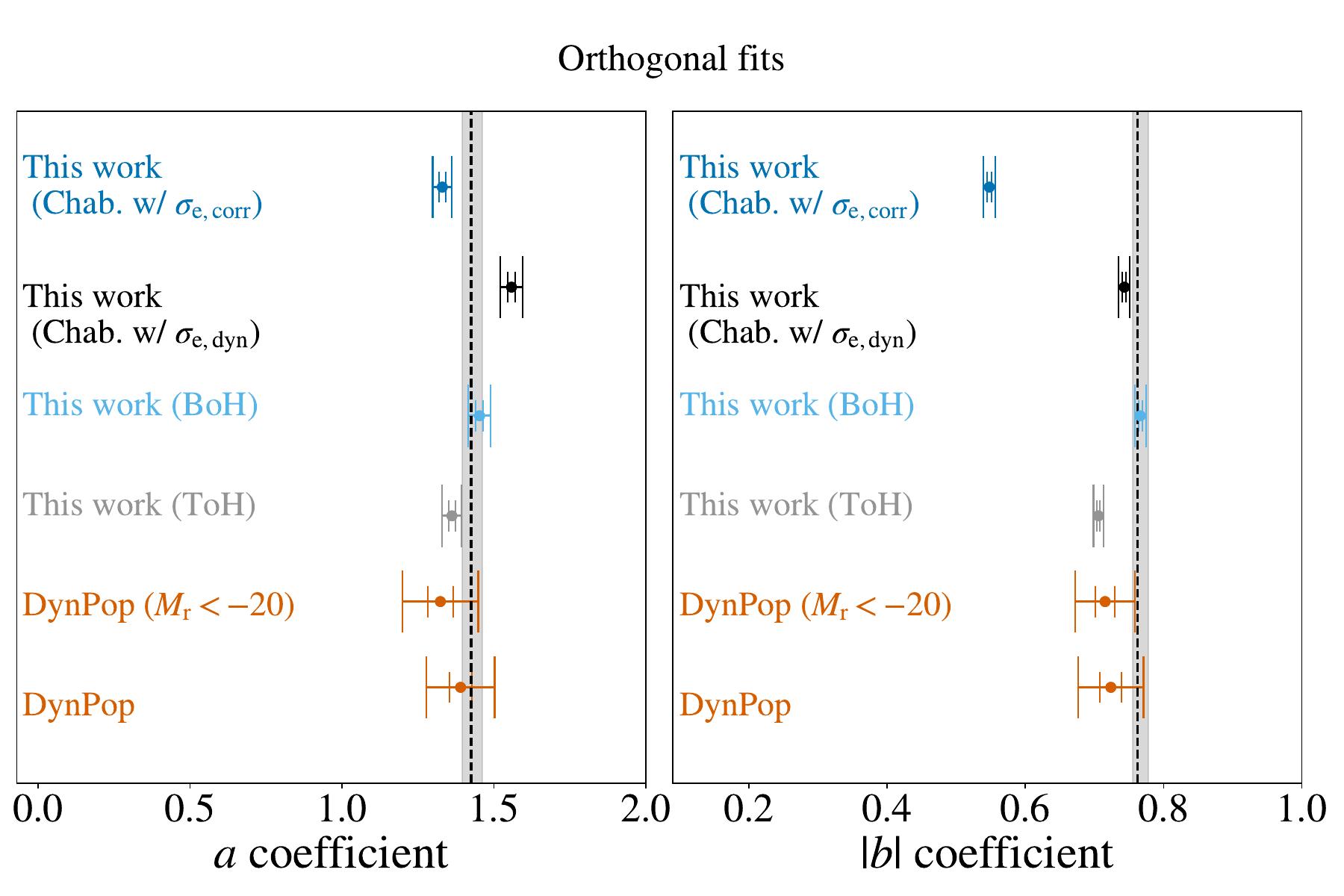}
 \caption{ {Best-fit coefficients of the fundamental plane from direct fit (left) and orthogonal fit (right) from our analysis compared to the DynPop catalogue and its luminosity-limited subsample ($M_{\rm r}< -20$). The small and larger error bars represent, respectively, the $1\sigma$ and $3\sigma$ confidence intervals of the inferences. As in the main text, the vertical dashed line represents the median of the observational inferences (except for DynPop), and the grey shaded region represents the 16th-84th percentiles.}}
 \label{fig:extended_data_fig_direct}
\end{figure*}

 {
\subsection*{Effect of the luminosity selection} In Figure \ref{fig:extended_data_fig_direct} 
%and \ref{fig:extended_data_fig_orthogonal} 
we examine how the FP coefficients vary when imposing a luminosity threshold of $M_{\rm r} < -20$, matching the limits of our virtual-ETG. Applying this cut to the DynPop sample, we illustrate that the absolute values of the FP coefficients become slightly shallower, consistent with the discussion in the main text. As with the correction estimated for the virtual-ETG in the main text, the impact of restricting the observational sample to $M_{\rm r} < -20$ on the FP coefficients remains within the measurement uncertainties.}

\begin{figure*}

    \centering
    \includegraphics[width=0.4\textwidth]{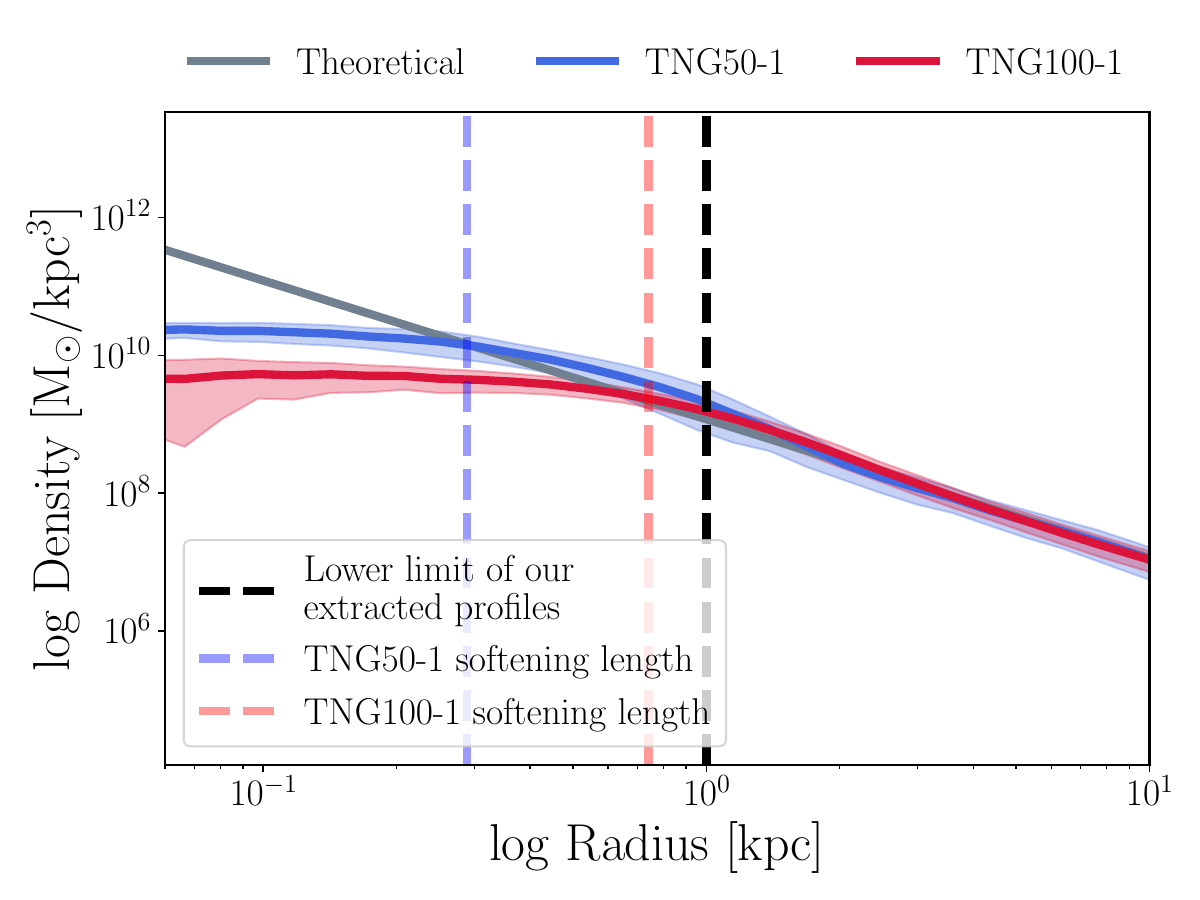}
    \hspace{+0.3cm}
        \includegraphics[width=0.4\textwidth]{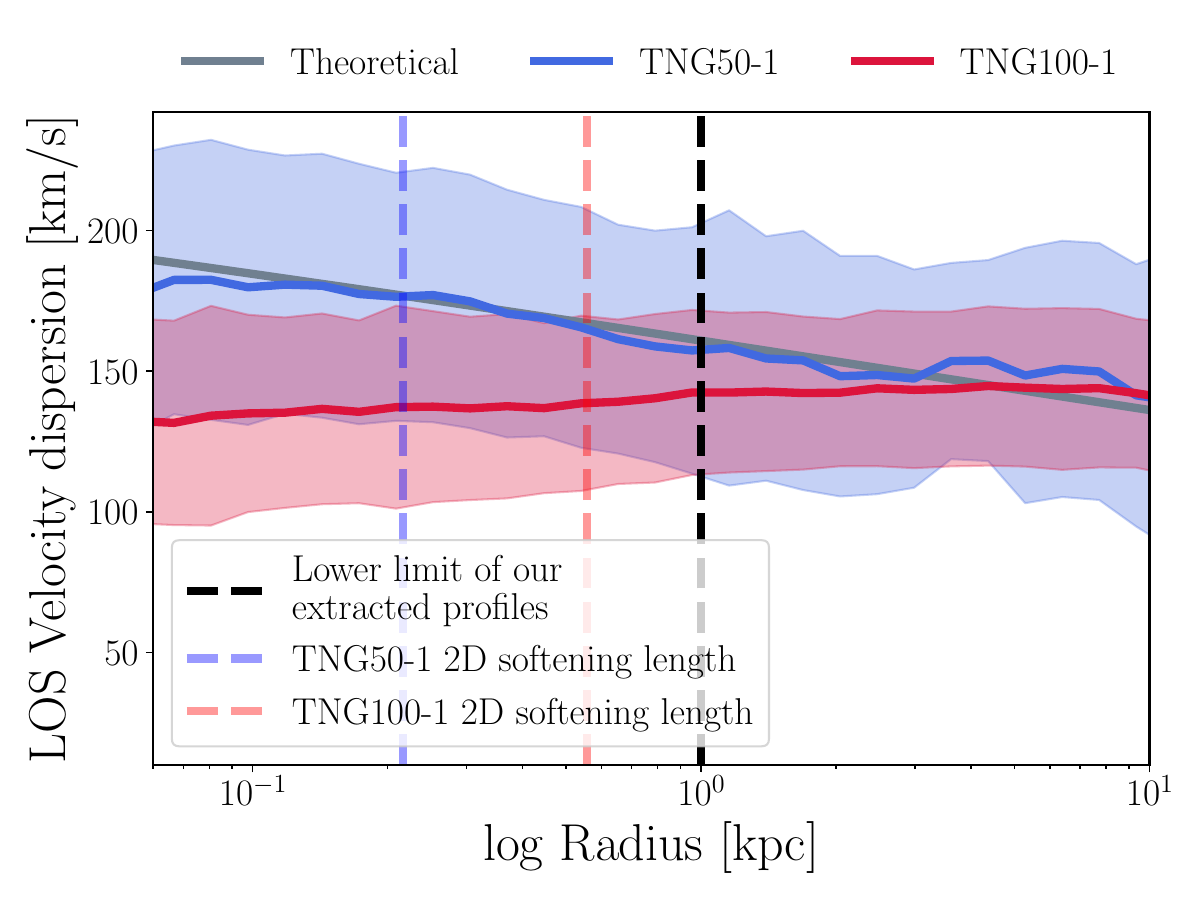}
    \caption{Resolution convergence between TNG50-1 and TNG100-1. We display their softening lengths as blue and red vertical dashed lines, respectively. For the LOS velocity dispersion profiles, the softening lengths are converted to 2D by dividing them by an approximate factor of $1.33$ \citep{Wolf}. For radii $r \gtrsim 1\,\mathrm{kpc}$ (black dashed line), both simulations yield consistent density and velocity dispersion profiles. These figures represent a stratified subsample of ETGs at $z=0$ ($\sim 370$ objects for each simulation), where the selection of TNG50-1 ETGs is analogous to the virtual-ETG, and they are defined over the sample stellar mass interval. The solid curves are the median trends of the density and velocity dispersion profiles, respectively, while the shaded regions represent the scatter around the median. The grey curves are the theoretical constructions of these median profiles using our modelling approach, i.e., a power-law density model to describe the matter distribution and the isotropic spherical Jeans equation described by this power-law density model and the de-projected S\'ersic profile.}
\label{fig:resolution_convergence_comparison}
\end{figure*}

 {
\subsection*{Resolution convergence test} In Figure \ref{fig:resolution_convergence_comparison}, We present the median density and velocity-dispersion profiles of our virtual ETGs from the TNG100-1 simulation, and compare them with those from the higher-resolution TNG50-1 run (softening length $~0.3$ kpc). The two simulations yield consistent profiles for $r \gtrsim 1$ kpc, indicating that our requirement that all constructed profiles be defined only above this radius leads to physically robust results. The innermost regions of the galaxies are then inferred through our analytical models, which include Sérsic-profile fitting, power-law density prescriptions, and solutions to the spherical Jeans equation. These profiles are defined between $0.05$ and $10$ kpc.} 

 {In the Figure mentioned above, we also show the analytical models (grey curves) specified by the median parameters of our virtual-ETG catalogue. As described in \S\ref{sec:Methods}, the median density profile is obtained by adopting a power-law form, and the line-of-sight (LOS) velocity dispersion profile is derived by solving the isotropic spherical Jeans equation, where the total mass profile follows this power-law density model and the tracer distribution is given by the de-projected S\'ersic profile.}

 {These results clearly show that our modelling accurately reproduces both the central density and LOS velocity dispersion profiles in regions beyond the simulation’s softening length. In particular, for the velocity dispersion, our method successfully recovers the innermost part of the profiles, in full agreement with the higher-resolution TNG50-1 simulation.}

%\begin{figure*}
%    \centering
%    \includegraphics[width=0.8\textwidth]{figures/luminosity_limited_dynpop_comparison_orthogonal.pdf}
%    \caption{ {Similar to Figure \ref{fig:extended_data_fig_direct}, but for orthogonal fits.}}
%    \label{fig:extended_data_fig_orthogonal}
%\end{figure*}

\section{Extended Fundamental Plane Data}
In Table \ref{tab:sigma_agreement}, we quantify the agreement between simulated and observed FPs by computing Z‐scores (absolute sigma deviations) for each study. Lower Z‐scores denote closer alignment, and boldface highlights the parameter choice that minimises discrepancy.

\begin{table}
\setcounter{table}{0}  % Set to desired number minus 1
%\small{\textbf{Extended Data Table 1. Z-scores between simulated and observed FPs.} Z-scores obtained between our FP measurements and other from the literature.}
\setlength{\abovecaptionskip}{0.1cm}
\setlength{\belowcaptionskip}{0.2cm}
\begin{center}

\caption{\label{tab:sigma_agreement}\textbf{Median Z-scores between simulated and observed FPs.} Lower values indicate better agreement. We present the results derived for the FP characterised by the three velocity dispersions introduced in the main text, as well as for the two fitting methods, i.e., direct (D) and orthogonal (O) fits.} 

\begin{tabular}{lcccccc}
\hline
Study                                     & Param. & Ch/$\sigma_{\rm e, corr}$ & Ch/$\sigma_{\rm e, dyn}$ & BoH              & ToH             & Fit                               \\ \hline
\multirow{2}{*}{Lu+20}         & $a$       & $9.52 $                        & $\mathbf{4.18 }$              & $6.74 $          & $8.14$          & \multirow{2}{*}{D}                         \\
                                          & $b$       & ${5.32 }$               & $6.70 $                       & $8.08 $          & $\mathbf{4.77}$          &                                                 \\
\multirow{2}{*}{B+03}   & $a$       & $3.27 $                        & $\mathbf{0.15 }$              & $1.53 $          & $2.42$          & \multirow{2}{*}{D}                         \\
                                          & $b$       & $18.2$                         & $3.93 $                       & $\mathbf{2.30 }$ & $6.31$          &                                                 \\
\multirow{2}{*}{H\&B09}  & $a$       & $2.45 $                        & $\mathbf{0.11}$               & $1.14 $          & $1.81$          & \multirow{2}{*}{D}                         \\
                                          & $b$       & $8.86 $                        & $2.15$                        & $\mathbf{1.39 }$ & $3.24$          &                                                 \\
\multirow{2}{*}{S+13}    & $a$       & $\mathbf{0.07 }$               & $4.46$                        & $2.31 $          & $1.16$          & \multirow{2}{*}{D}                         \\
                                          & $b$       & $14.49 $                       & $3.29$                        & $\mathbf{2.02 }$ & $5.13$          &                                                 \\
\multirow{2}{*}{B+20}   & $a$       & $11.41 $                       & $\mathbf{3.81}$               & $7.39$           & $9.38$          & \multirow{2}{*}{D}                         \\
                                          & $b$       & $8.97 $                        & ${3.13 }$                     & $4.50 $          & $\mathbf{1.17}$ &                                                 \\
\multirow{2}{*}{Y\&P20}      & $a$       & $\mathbf{2.53}$                & $13.47$                       & $8.28$           & $5.40$          & \multirow{2}{*}{D}                         \\
                                          & $b$       & $42.92$                        & $6.00$                        & $\mathbf{1.71}$  & $12.63$         &                                                 \\ \hline
\multirow{2}{*}{B+03}   & $a$       & $3.13 $                        & $1.32 $                       & $\mathbf{0.72 }$ & $2.51$          & \multirow{2}{*}{O}                     \\
                                          & $b$       & $19.47 $                       & $\mathbf{0.65} $              & $1.57 $          & $4.31$          &                                                 \\
\multirow{2}{*}{LB+08} & $a$       & $1.76 $                        & $2.68$                        & $\mathbf{0.64}$  & ${1.14}$        & {\multirow{2}{*}{O}} \\
                                          & $b$       & $25.03 $                       & $1.98 $                       & $\mathbf{0.74 }$ & $6.50$          & \multicolumn{1}{l}{}                            \\
\multirow{2}{*}{H\&B09}  & $a$       & $4.42 $                        & $5.44 $                       & $\mathbf{0.98 }$ & ${3.00}$        & {\multirow{2}{*}{O}} \\
                                          & $b$       & $11.99 $                       & $2.31$                        & $\mathbf{1.18 }$ & $4.18$          & \multicolumn{1}{l}{}                            \\
\multirow{2}{*}{H\&W15}       & $a$       & ${3.81} $                      & $14.26 $                      & $6.23 $          & $\mathbf{0.97}$ & \multirow{2}{*}{O}                     \\
                                          & $b$       & $72.86$                        & $7.17 $                       & $\mathbf{0.78 }$ & $22.55$         &                                                 \\ \hline
\end{tabular}

\end{center}

Note: the result from \cite{Lu+2020} is the only one that comes from simulated data, but is reported in the Table for completeness. Abbreviations: Lu+20 \citep{Lu+2020}; B+03 \citep{Bernardi+2003}; H\&B09 \citep{Hyde+2009}; S+13 \citep{Saulder+2013}; B+20 \citep{Bernardi+2020}; Y\&P20 \citep{Yoon_2020}; LB+08 \citep{LaBarbera+2008}; H\&W15 \citep{Hou_2015}.
\end{table}
\setcounter{table}{1}
%\begin{figure*} 
%    \begin{center}
%    \includegraphics[width=2\columnwidth]{figures/edge_on_projections_logsigma_rm_e,_delta_IMF.pdf}\\
%    \end{center}
%    \small{\textbf{Extended Data Figure 1. The virtual-ETG FP fitted with a velocity dispersion rescaled to take into account IMF variations.} We show both direct (left panel) and orthogonal (right panel) fits. For the former, the $b$ coefficient is smaller than the median value obtained from observational works and for the latter the $a$ coefficient is larger than the median value obtained from observational works.}
    %\label{fig:supplementary_fig_sigma_dyn}
%\end{figure*}

%\renewcommand{\figurename}{Extended Data Figure}
%\setcounter{figure}{0}  
%\renewcommand{\figurename}{Figure}

%%%%%%%%%%%%%%%%%%%%%%%%%%%%%%%%%%%%%%%%%%%%%%%%%%

\section{Two-dimensional scaling relations}\label{appendix:2d_scaling_relations}

In this Appendix, we introduce some scaling relations used for interpreting the virtual-ETG FP tilt in terms of homology and M/L trends (see \S\ref{sec:virial_consistency_and_feedback}).

\begin{figure}
    \centering
    \includegraphics[width=0.98\columnwidth]{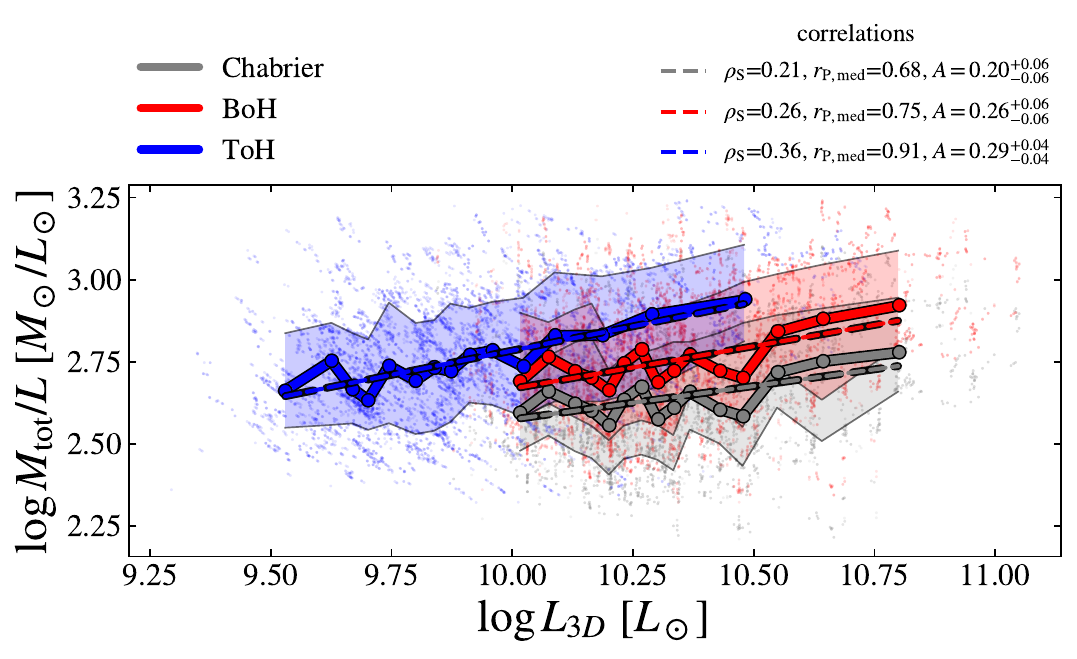}
    \includegraphics[width=0.98\columnwidth]{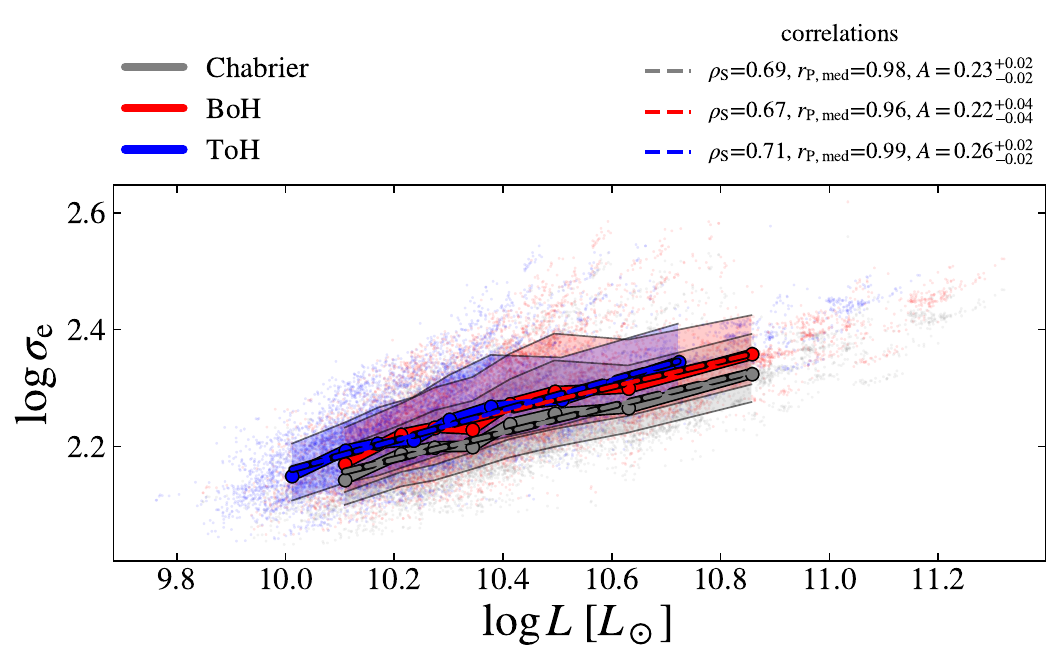}
    \includegraphics[width=0.98\columnwidth]{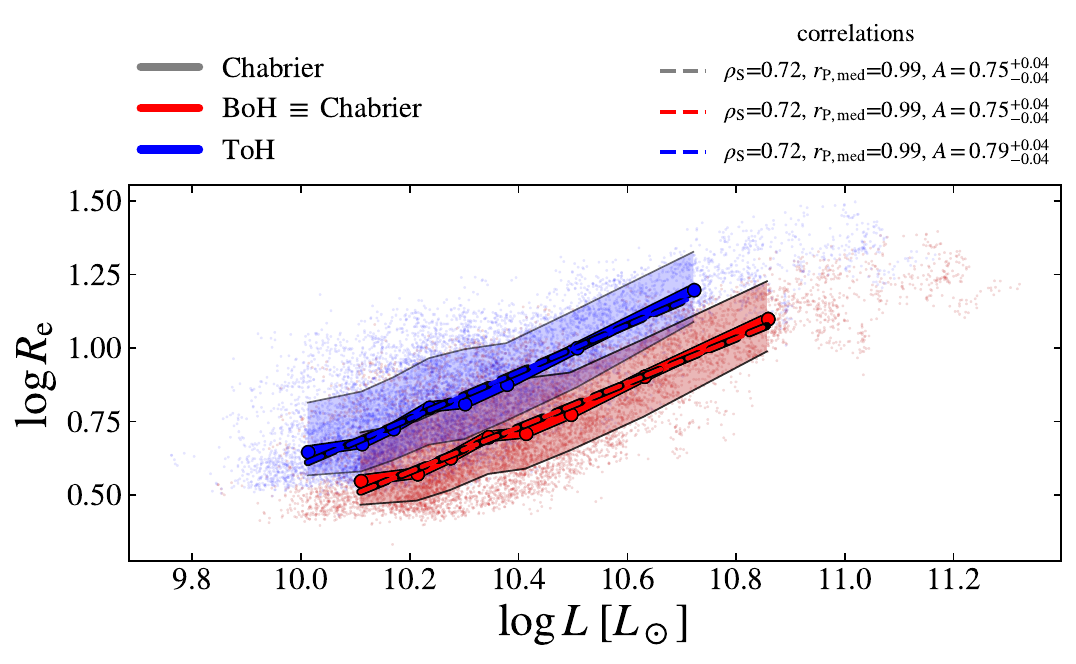}
    \caption{Two-dimensional scaling relations of the virtual ETGs for all IMF models examined in this work. In each panel, we display the corresponding IMF indicated by different line colors, along with their summary statistics. We report the Spearman rank correlation coefficient, $\rho_{\rm s}$, which quantifies the strength of non-linear correlations between the variables: values approaching $1$ ($-1$) indicate a strong positive (negative) monotonic correlation. We also present the Pearson correlation coefficient, $r_{\rm p}$, for which, analogous to the Spearman coefficient, values close to $1$ ($-1$) denote a strong positive (negative) linear correlation. Finally, we list the best-fit slopes ($A$) for each line in the diagrams.}
    \label{fig:2D_scaling_relations}
\end{figure}

% Don't change these lines
\bsp	% typesetting comment
\label{lastpage}
\end{document}